\documentclass[twocolumn,twocolappendix]{aastex6}		

\newcommand{\myemail}{pzeidler@ari.uni-heidelberg.de}
\usepackage{natbib}

\shorttitle{The PDMF of Westerlund~2}
\shortauthors{Zeidler et al.}

\begin{document}

\title{A High-Resolution Multiband Survey of Westerlund 2 With the \textit{Hubble Space Telescope}. III. The present-day stellar mass function}

\author{Peter Zeidler\altaffilmark{1,2}, Antonella Nota\altaffilmark{2,3}, Eva K. Grebel\altaffilmark{1}, Elena Sabbi\altaffilmark{2}, Anna Pasquali\altaffilmark{1}, Monica Tosi\altaffilmark{4}, \and Carol Christian\altaffilmark{2}}

\altaffiltext{1}{Astronomisches Rechen-Institut, Zentrum f\"ur Astronomie der Universit\"at Heidelberg, M\"onchhofstr. 12-14, 69120 Heidelberg, Germany \email{\myemail}}
\altaffiltext{2}{Space Telescope Science Institute, 3700 San Martin Drive, Baltimore, MD 21218, USA}
\altaffiltext{3}{ESA, SRE Operations Devision, Spain}
\altaffiltext{4}{INAF - Osservatorio Astronomico di Bologna}

\begin{abstract}
We present a detailed analysis of the spatial distribution of the stellar population and the present-day mass function (PDMF) of the Westerlund~2 (Wd2) region using the data from our high resolution multi-band survey with the \textit{Hubble} Space Telescope. We used state-of-the-art artificial star tests to determine spatially resolved completeness maps for each of the broad-band filters. We reach a level of completeness of 50\% down to $F555W=24.8$~mag ($0.7~M_\odot$) and $F814W=23.3$~mag ($0.2~M_\odot$) in the optical and $F125W=20.2$~mag and $F160W=19.4$~mag (both $0.12~M_\odot$) in the infrared throughout the field of view. We had previously reported that the core of Wd2 consists of two clumps: namely the main cluster (MC) and the northern clump (NC). From the spatial distribution of the  completeness corrected population, we find that their stellar surface densities are $1114~\rm{stars~pc}^{-2}$ and $555~\rm{stars~pc}^{-2}$, respectively, down to $F814W=21.8$~mag. We find that the present-day mass function (PDMF) of Wd2 has a slope of $\Gamma=-1.46 \pm 0.06$, which translates to a total stellar cluster mass of $(3.6 \pm 0.3)\cdot 10^4~M_\odot$. The spatial analysis of the PDMF reveals that the cluster population is mass-segregated, most likely primordial. In addition, we report the detection of a stellar population of spatially uniformly distributed low-mass ($<0.15~M_\odot$) stars, extending into the gas ridges of the surrounding gas and dust cloud, as well as a confined region of reddened stars, likely caused by a foreground CO cloud. We find hints that a cloud-cloud collision might be the origin of the formation of Wd2.

\end{abstract}


\section{Introduction}
\label{sec:introduction}
With an estimated stellar mass of M~$\ge 10^4$~M$_\odot$ \citep{Ascenso_07}, the young Galactic star cluster \object{Westerlund~2} \citep[][; hereafter Wd2]{Westerlund_61} is one of the most massive young clusters in the Milky Way (MW). Located in the Carina-Sagittarius spiral arm at $(\alpha,\delta)=(10^h23^m58^s.1,-57^\circ45'49'')$(J2000), $(l,b)=(284.3^\circ,-0.34^\circ)$, Wd2 is the central cluster of the \ion{H}{2} region \object{RCW~49} \citep{Rodgers_60}.

In our first paper \citep[][hereafter Paper~I]{Zeidler_15}, we introduced our multi-band survey with the \textit{Hubble} Space Telescope (HST), presented the data reduction, and confirmed the cluster distance estimate by \citet{Vargas_Alvarez_13} of 4.16~kpc based on a spectroscopic and photometric analysis of Wd2's stellar population. In \citet[][hereafter Paper~II]{Zeidler_16b} we estimated a mean age of $1.04\pm0.72$~Myr for the cluster region with a possibility for a small age gradient of 0.15~Myr from the center to the surrounding molecular cloud. The age estimate is in good agreement with the expected lifetimes of the O3--O5 main-sequence (MS) stars of $\sim2-5$~Myr \citep[see Tab.~1.1,][]{Sparke_07}. Such stars were spectroscopically identified in the center of Wd2 \citep{Rauw_07,Rauw_11,Vargas_Alvarez_13}. Using two-color-diagrams (TCDs), we found a total-to-selective extinction of $R_V=3.95 \pm 0.135$. This value was confirmed by an independent numerical study of \citet{Mohr-Smith_15}. Their best-fitting parameter is $R_V=3.96^{+0.12}_{-0.14}$, which is in very good agreement with our result. We also showed that the Wd2 cluster consists of two subclusters, namely the main cluster (MC) and the northern clump (NC).

The analysis of the spatial distribution of stars in young massive star clusters is challenging due to variable extinction caused by gas and dust, and due to the crowding caused by the high stellar density. Both gas and dust, as well as the high stellar density, result in a spatially dependent limited observational depth and completeness. In regions with low gas and dust content, and a lower stellar surface density, even the youngest and faintest stars can be resolved, while in regions such as the cluster center or in close proximity to luminous O and B-stars, many of the low-mass stars are lost due to crowding. The common approach to quantify these completeness effects is to run artificial star tests \citep[see, e.g.,][]{Stolte_05,Pang_13,Cignoni_15,Sabbi_16} and to statistically correct the number of stars per luminosity and spatial bin.

In this paper we present the artificial star tests for all four bands of our multi-band HST survey (Paper~I) and the related spatially resolved completeness maps. We use these maps to analyze the spatial distribution of the rich population of pre-main-sequence (PMS) stars of Wd2. In particular, we focus on the two density clumps in the cluster already identified by \citet{Vargas_Alvarez_13}, \citet{Hur_15}, and \citet{Zeidler_15}.

The mass function (MF) of a system provides the distribution of stars as a function of mass. Understanding this distribution of mass is important for the formation and evolution of neighboring stars, star clusters, and galaxies, since it influences the dynamics and evolution of such systems, as well as the properties of the next generations of stars. There is on-going debate whether or not the MF is universal (e.g., with a \citet{Salpeter_55} or \citet{Kroupa_01}-like high-mass slope of $\Gamma \sim -1.3$) or whether it varies in different environments and is a property of each individual system. Our deep observations allow us to study the present-day mass function (PDMF) down to $\sim 0.5 M_\odot$. The only earlier study of the PDMF of Wd2 was performed by \citet{Ascenso_07}, based on ground-based NIR observations ($J$, $H$, $K_s$). \citet{Ascenso_07} derived a slope of $\Gamma=-1.20 \pm 0.16$.

A study by \citet{Weisz_15} in M31 as part of the Panchromatic Hubble Andromeda Treasury program \citep[PHAT,][]{Dalcanton_12} shows a quite universal slope of $\Gamma=-1.45^{+0.03}_{-0.06}$ for the high-mass PDMF ($m \ge 1~\rm{M}_\odot$), for resolved stars in young clusters. Studies in the MW show a larger scatter between different massive clusters \citep[$\sigma \sim $0.3--0.4, e.g.,][]{Weisz_15,Massey_03}, with, e.g., $\Gamma=-0.88 \pm 0.15$ for NGC~3603 \citep{Pang_13}, $\Gamma=-0.8 \pm 0.15$ for the Arches cluster \citep{Stolte_02}, and $\Gamma=-1.45 \pm 0.12$ for NGC~6611 \citep{Bonatto_06}.

Young star clusters such as in the Orion Nebula Cluster \citep[ONC,][]{Hillenbrand_98}, the Arches cluster \citep{Stolte_02}, or NGC~3603 \citep{Pang_13} show mass segregation, yet the origin is still unclear. Due to the young age of the Trapezium cluster in the ONC \citet{Bonnell_98} argued that the massive stars did not yet have the time to move to the cluster center but were born at their current location. The primordial mass segregation scenario is supported by the competitive accretion theory of \citet{Bonnell_01} and \citet{Bonnell_06}, which suggests that protostars can accrete more mass in the dense central regions of a young star cluster than in the outer regions. On the other hand, \citet{McMillan_07} argue that young clusters showing mass segregation may be a result of mergers between small clumps that are already dynamically mass-segregated. \citet{Krumholz_09} and \citet{Moeckel_09b} argue that mass segregation can even occur on very short time scales so that the massive stars can be formed anywhere in the cluster.

This paper is a continuation study based on the observations presented in Paper~I. In Paper~II we presented the examination of the PMS stars in Wd2 and the effects of the OB star population on the protoplanetary disk evolution. We used the H$\alpha$ line emission to select mass-accreting PMS stars with H$\alpha$ excess emission. Following the method outlined by \citet{deMarchi_10}, a total of 240 stars with active mass accretion were selected. We showed that the mass accretion rate tend to be lower in close proximity to the massive OB-stars, whose far-ultra-violet flux may cause an accelerated dispersal of their circumstellar disks.

The current paper focuses on the spatial distribution of the stellar population in RCW~49 and the PDMF of the Wd2 cluster. In Sect.~\ref{sec:catalog} we give a brief summary of our observations and the photometric catalog as well as the main results from Paper~I and II. In Sect.~\ref{sec:completeness} we present the performed artificial star tests and the creation of the completeness maps for the different filters. In Sect.~\ref{sec:dist}  we focus on the distribution of the stellar population of RCW~49. We also discuss the scenario of a cloud-cloud collision as the possible origin of Wd2. In Sect.~\ref{sec:mass_function} we determine and discuss the present-day stellar MF of Wd2 as well as mass segregation. In Sect.~\ref{sec:summary} we present a summary of the paper as well as our conclusions.

\section{A brief overview of the observations and recent work}
\label{sec:catalog}
Our observations of Wd2 were performed with the \textit{Hubble} Space Telescope (HST) during Cycle 20 using the Advanced Camera for Surveys \citep[ACS,][]{ACS} and the Wide Field Camera 3 in the IR channel \citep[WFC3/IR,][]{WFC3}. In total, six orbits were granted and the science images were taken on 2013 September 2 to 8 (proposal ID: 13038, PI: A. Nota). For a detailed description, we refer to Paper~I.

Observations in four wide-band filters ($F555W$: $\lambda_P=5361.0$~\AA, $\Delta \lambda=360.1$~\AA ; $F814W$: $\lambda_P=8057.0$~\AA, $\Delta \lambda=652.0$~\AA , $F125W$: $\lambda_P=12,486.0$~\AA, $\Delta \lambda=866.28$~\AA , and $F160W$: $\lambda_P=15,369.0$~\AA, $\Delta \lambda=826.28$~\AA ) were obtained. Imaging in two narrow-band filters, $F658N$ ($\lambda_P=6584.0$~\AA, $\Delta \lambda=36.8$~\AA) and $F128N$ ($\lambda_P=12,832.0$~\AA, $\Delta \lambda=357.45$~\AA), centered on the H$\alpha$ and Pa$\beta$ line emissions, respectively, was added to the wide-band observations. The final photometric catalog contains 17,121 point sources in total with a two-band-detection minimum requirement. 2236 point sources were detected in all six filters while 3038 point sources were detected in all four wide-band filters.

The combined information from the $F658N$ and $F128N$ filters allowed us to create a high-resolution pixel-to-pixel (0.098 arcsec pixel$^{-1}$) $E(B-V)_g$ color excess map of the gas. This map was then used to individually deredden our photometric catalog (see Paper~I).  All colors and magnitudes denoted with the subscript '0' were individually dereddened using our pixel-to-pixel color excess map.

Throughout this paper, if not stated differently, we will use $d=4.16$~kpc and $R_V=3.95 \pm 0.135$ \citep{Vargas_Alvarez_13,Mohr-Smith_15,Zeidler_15} for Wd2.

\section{Completeness Correction}
\label{sec:completeness}

The high number of luminous OB stars, the high stellar surface density, and the differential reddening across the Wd2 area cause varying position-dependent detection limits for fainter stars. In the cluster center, where crowding is significant, the brightness limit for a certain detection fraction is, as expected, higher than in the less densely populated outer regions. This impacts the estimate of the stellar surface density and density profile, especially since completeness depends on location and filter.

To characterize and compensate for these completeness effects a common and often applied technique is used: artificial star tests. This method uses a dense grid of artificially created stars throughout the field of view (FOV). It adds artificial stars to the image, which are then retrieved using the same process applied to real data. In order to provide a statistically significant correction factor, the artificial star tests were performed $5 \cdot 10^6$ times. Each time, only one artificial star was added in order to minimize an increase of crowding. The input catalog is well controlled in terms of magnitude and position of the input star, so it is possible to determine the exact fraction of artificial stars recovered for a certain brightness and position. The high surface density of the artificial stars ensures that the FOV is well covered and despite a random but uniform grid of the locations of the artificial stars, that all pixels of the observed images are well covered.

\subsection{The artificial star test}
\label{sec:artificial_test}

\begin{deluxetable}{lrrr}
	\tablecaption{Artificial star test results\label{tab:artificial_test}}
	\tablehead{
		\multicolumn{1}{c}{ } &\multicolumn{2}{c}{\textsc{number of artificial stars}} & \multicolumn{1}{c}{$\Sigma$}\\
		\multicolumn{1}{c}{\textsc{Filter}} &\multicolumn{1}{c}{\textsc{input}} &\multicolumn{1}{c}{\textsc{recovered}}  &\multicolumn{1}{c}{\textsc{$[\rm{px}^{-1}]$}}
	}
	\startdata
	$F555W$	&	1\,847\,822  & 1\,446\,659 & 0.074 \\
	$F814W$	&	1\,847\,822  & 1\,496\,963 &  0.077 \\
	$F125W$	&	1\,960\,996  & 1\,575\,256 &  0.075 \\
	$F160W$	&	1\,960\,996  & 1\,575\,289 &  0.075 \\
	\enddata
	\tablecomments{In this table we present the result of the artificial star tests. The first column shows the filter. Column~2 shows the number of artificial stars inside the FOV of each stacked and mosaiced observation. Column~3 shows the number of recovered artificial stars, applying the same cuts as in Paper~I for the scientific photometric catalog. Column~4 shows the mean stellar surface density per pixel.}
\end{deluxetable}

A grid of $5 \cdot 10^6$ artificial stars was created and used for all filters. The artificial star catalogs are populated down to $\sim30$~mag and are at least 2~mag deeper than the observed photometric catalog to ensure a proper coverage of the magnitude and colors of the stars. The color dependence is important because the stars used to determine the MF need to be detected in two filters simultaneously, e.g., an IR-bright star can be too faint in the optical to be detected. Therefore, this star needs to be counted as lost when deriving the surface density profile or the MF. We used the observed color of the stars to retrieve the completeness fraction of each observed star at its spatial position for each filter combination (see eq.~\ref{eq:completness_mult_bands}), depending on which filter combination one uses. As a result, we obtained the completeness information at each position of the survey area per magnitude and color bin (dependent on the filter combination), thus we can correct each star individually (depending on its position, magnitude, and color). The size of the artificial star grid was chosen in a way that each bin contains an average number of 500 artificial stars (bin size is: $\sim 91~\rm{px}^2$ or $\sim 0.22~\rm{arcsec}^2$). The magnitude binsize is 0.1~mag. We now can assign to each star in each of the wide-band filters its individual completeness level.

For the artificial star tests we used the photometry software DOLPHOT\footnote{\url{http://americano.dolphinsim.com/dolphot/}} \citep{Dolphin_00}, with the same setup as adopted for the original photometry. To clean the catalogs from spurious sources we applied the same statistical cuts used in Paper~I. Additionally, those artificial stars that coincided with very bright stars and hence had been recovered at least magnitudes 0.75~mag brighter\footnote{A recovered magnitude 0.75~mag brighter than the input magnitude means that the majority of the flux is emitted by surrounding stars due to crowding. 0.75~mag is the threshold where one measures at least twice the input flux $0.75=2.5\cdot \log_{10} (2)$} than their input magnitudes, were treated as lost. Real objects that faint located close to a bright star cannot be detected since we would detect only the bright source. In Tab.~\ref{tab:artificial_test} we present the numbers of the artificial star tests for each filter before and after the cleaning process. In column~2 we provide the input number of artificial stars within the FOV and in column~3 the total number of recovered stars after all statistical cuts were applied.

\subsection{The global completeness limits of Wd2}
\label{sec:global_completeness}

\begin{figure*}[htb]
	\resizebox{\hsize}{!}{\includegraphics{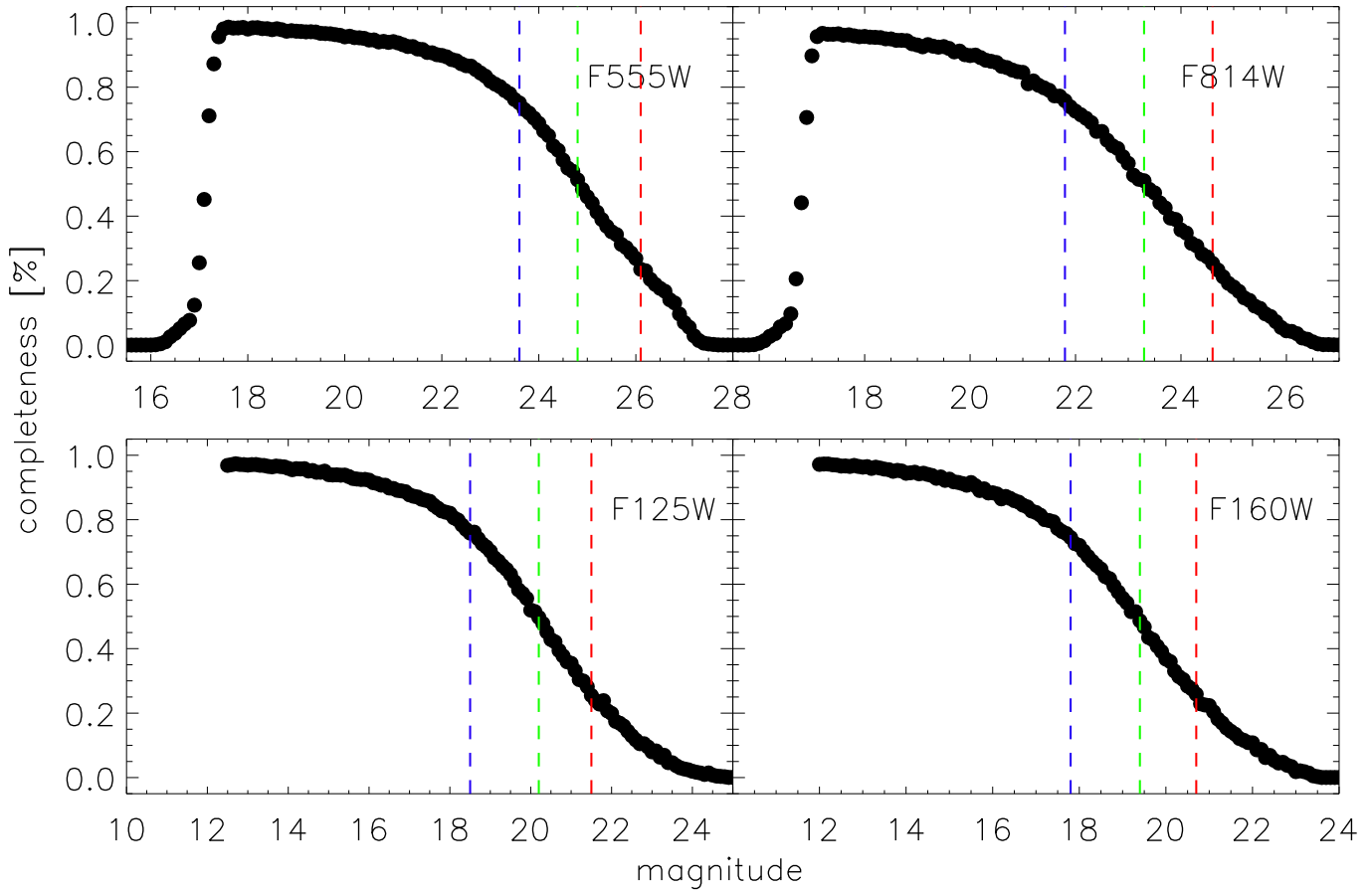}}
	\caption{The mean completeness distribution for the four wide band filters with a binsize of 0.1~mag. The dashed lines represent the 75\% (blue), 50\% (green), and 25\% (red) completeness limits (see Tab.~\ref{tab:completeness}).}
	\label{fig:completeness}
\end{figure*}

The level of completeness is defined by the ratio between the number of recovered stars and the number of input stars for each magnitude bin and position for a binsize of 0.1~mag. The completeness level as a function of magnitude for the four wide-band filters is presented in Fig.~\ref{fig:completeness}. We reach a completeness level of 50\% (green dashed lines in Fig.~\ref{fig:completeness}) at the following magnitudes: $F555W=24.8$~mag, $F814W=23.3$~mag, $F125W=20.2$~mag, and $F160W=19.4$~mag. The 75\% limit (blue dashed lines in Fig.~\ref{fig:completeness})  is reached at the following magnitudes: $F555W=23.6$~mag, $F814W=21.8$~mag, $F125W=18.5$~mag, and $F160W=17.8$~mag. An overview is given in Tab.~\ref{tab:completeness}.

Toward the bright end, the completeness of the ACS filters decreases drastically. This is the point at which the saturation limit of the detector is reached. For the ACS filters we recover the brighter stars using the 3~s short exposures to complete the catalog. Yet, the brightest stars are still saturated.

\subsection{The 2D completeness map of Wd2}
\label{sec:completeness_maps}

The spatially variable extinction and position-dependent surface density require position-dependent completeness maps. In this way the stellar numbers can be corrected as a function of spatial position in the image.

The completeness fractions can be defined as the probability of detecting a star of certain magnitude in a specific region of the observation. Therefore, to detect a star in multiple bands, or in other words to take into account the color of the star, the completeness fractions are multiplicative at each individual position in the FOV:

\begin{equation}
\label{eq:completness_mult_bands}
N_{\rm{corr.}}=\prod_{n=1}^{\#_{\rm{filter}}} \frac{1}{f_n} \cdot N_{\rm{obs.}},
\end{equation}

where $N$ is the number of observed and completeness-corrected stars and $f_{\rm n}$ completeness fraction at each position.
 
\begin{figure*}[htb]
	\begin{minipage}{\textwidth}
	\plottwo{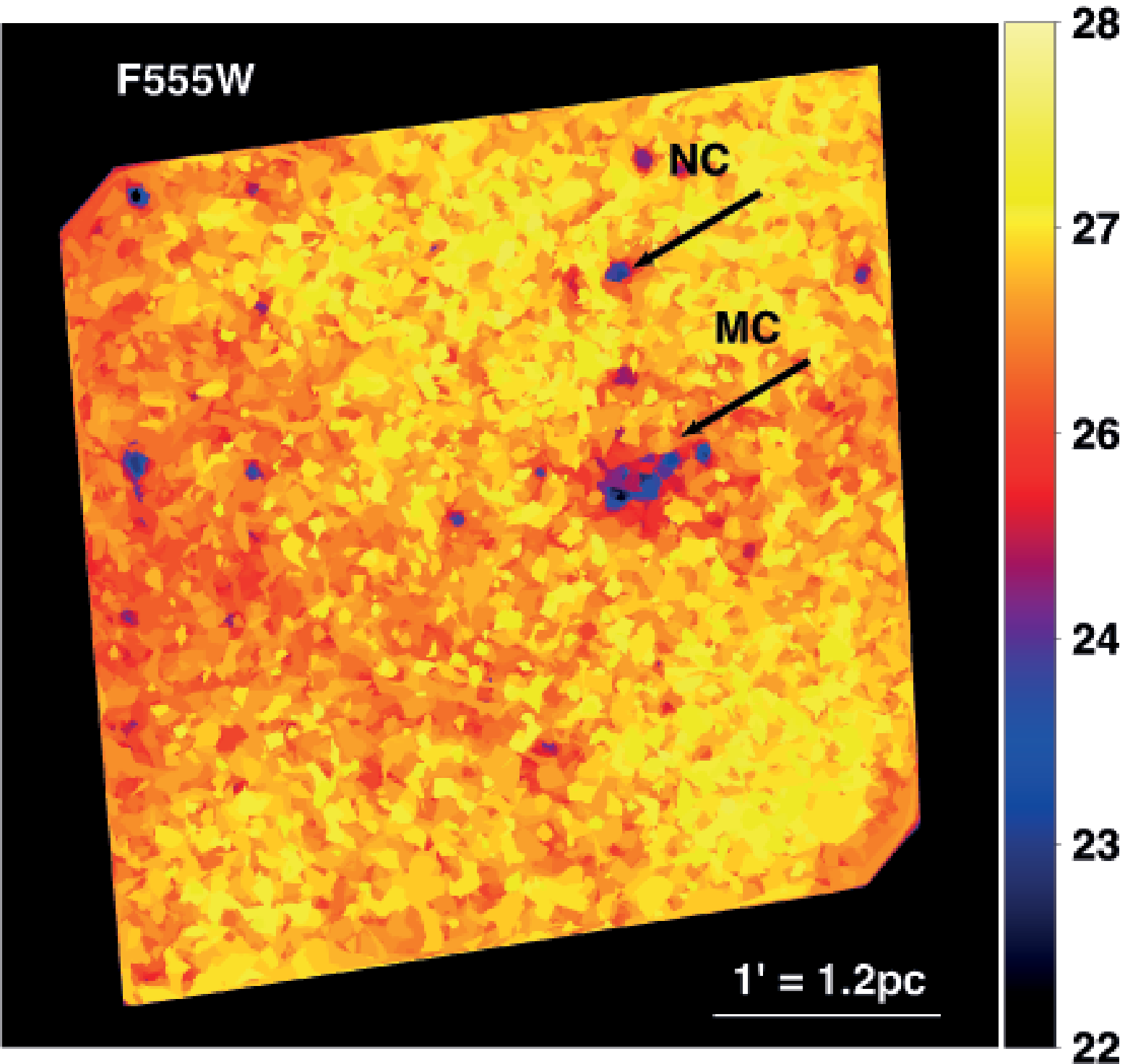}{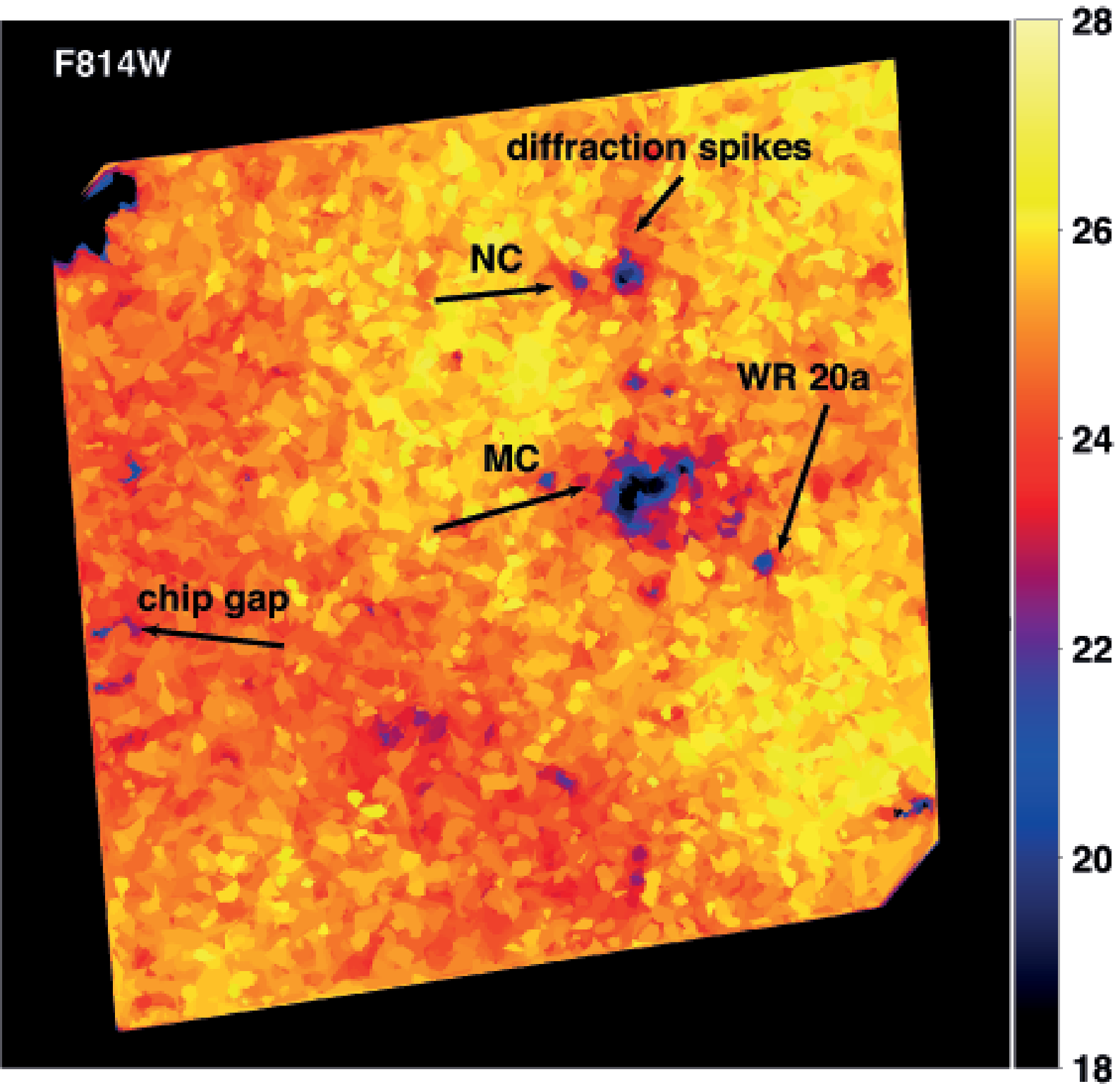}
	\end{minipage}
	\begin{minipage}{\textwidth}
		\plottwo{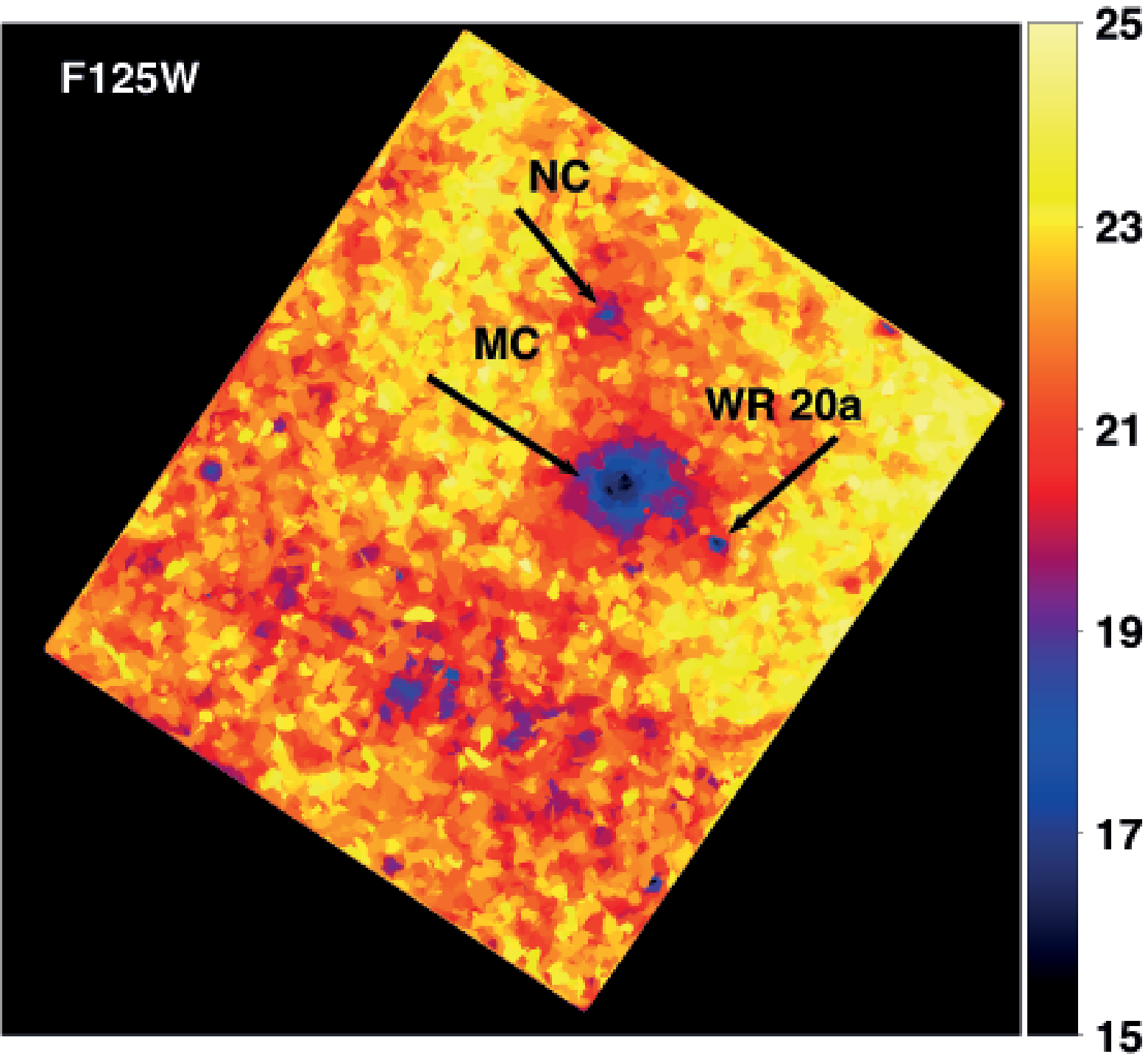}{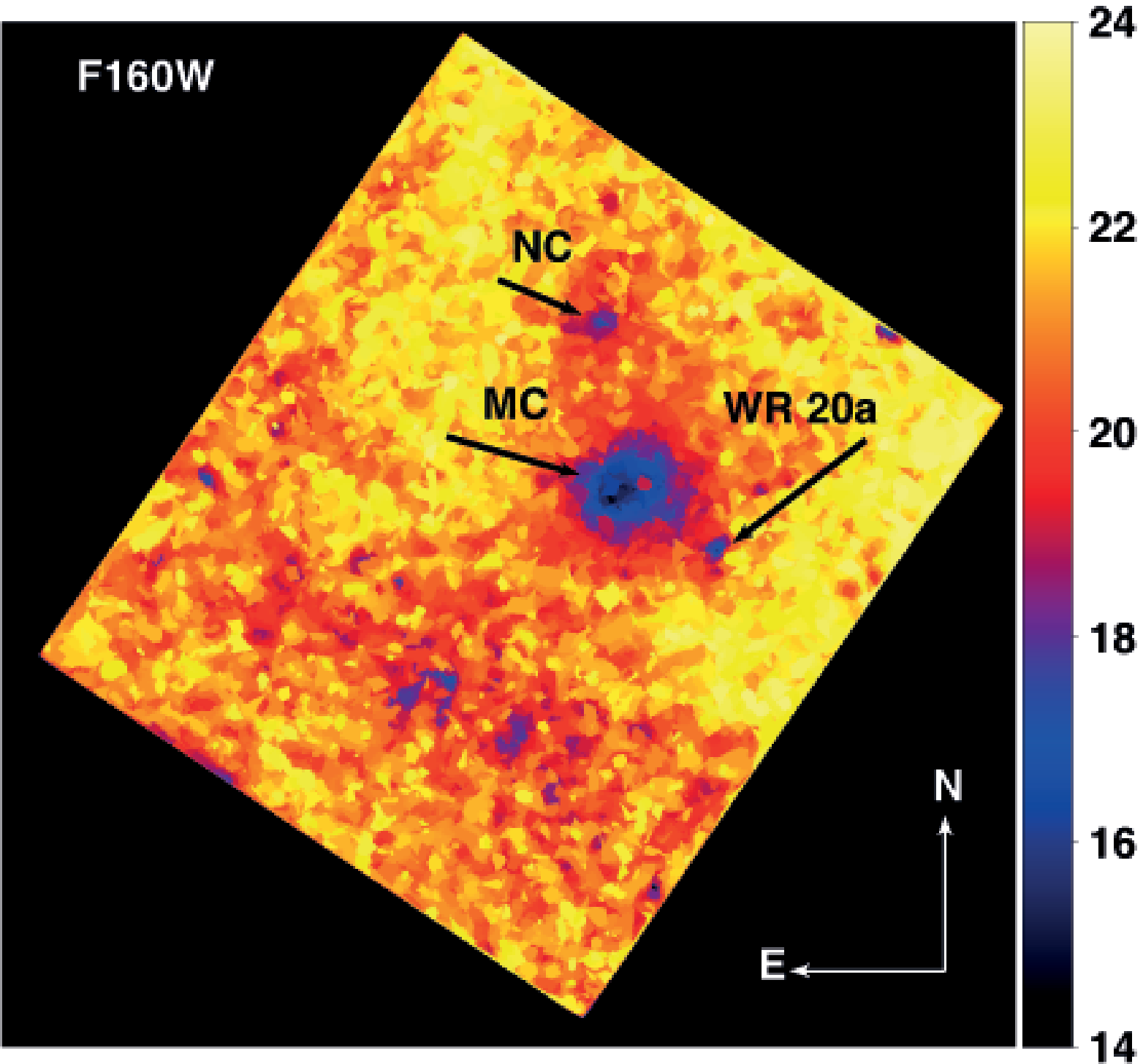}
	\end{minipage}
\caption{The 50\% completeness map for all four wide-band filters. The color bar indicates the magnitude at which such a completeness value is reached. We marked some features in each of the panels, such as the two clumps (MC, NC), the Wolf-Rayet star WR20a, the imprint of diffraction spikes, and the defect in the image caused by the chip gaps for the ACS camera.}
	\label{fig:completeness_50}
\end{figure*}

In Fig.~\ref{fig:completeness_50} we show the 2D maps for a completeness limit of 50\% in the four filters. The different colors indicate the corresponding magnitude limit. In Tab.~\ref{tab:completeness} we list the mean magnitude for a completeness limit in several regions including the gas and dust ridge. The masses are computed using a 1~Myr isochrone at Solar metallicity of $Z_\odot=0.0152$ taken from the \textsc{PAdova and TRieste Stellar Evolution Code}\footnote{\url{http://stev.oapd.inaf.it/cmd}} \citep[hereafter: PARSEC 1.2S,][]{Bressan_12}.

Besides the two clumps and several luminous stars, such as foreground stars, the luminous Wolf-Rayet star WR20a \citep[e.g.,][]{Bonanos_04}, and their diffraction spikes, the 50\% completeness level is quite uniformly distributed throughout the whole FOV in all filters (see Fig.~\ref{fig:completeness_50}). While in the $F555W$ filter stars of masses of $0.7~M_\odot$ can be detected (0.41--0.8$~M_\odot$ in the two clumps, depending on the position), $0.2~M_\odot$ stars can only be detected in the $F814W$ filter (0.12--0.6$~M_\odot$ in the two clumps, depending on the position). In the IR filters stars with a mass of $0.1~M_\odot$ or lower are detected (0.44--0.75$~M_\odot$ in the two clumps, depending on the position). We emphasize that these numbers are mean values across different regions. 

\subsection{The true photometric uncertainties}
\label{sec:true_phot_uncert}
\begin{figure*}[htb]
	\resizebox{0.95\hsize}{!}{\includegraphics{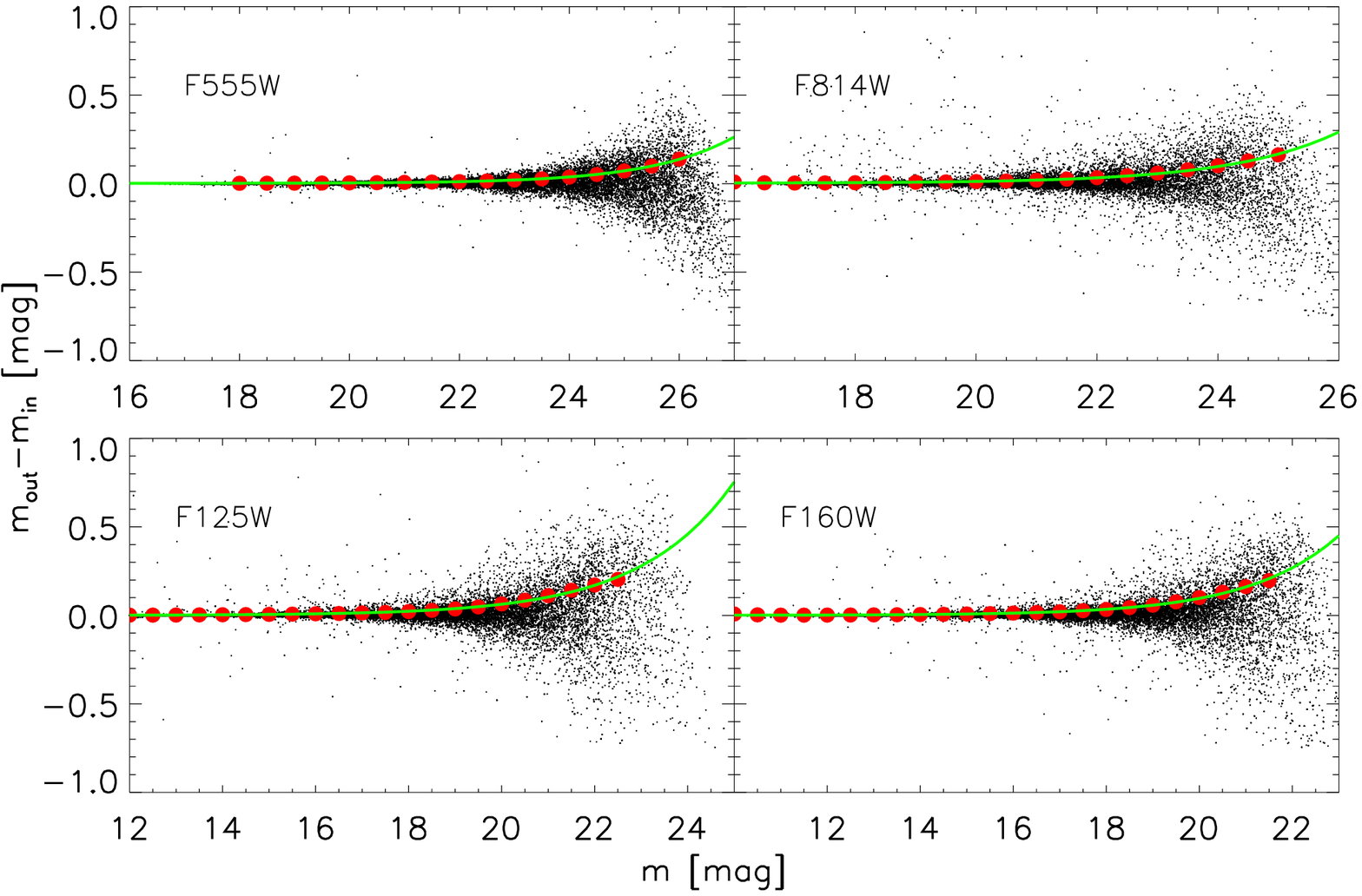}}
	\caption{The difference between the input magnitude and the measured magnitude of the artificial stars as a function of magnitude. The green line is an exponential fit of the absolute mean scatter and defines the true photometric uncertainties. For visualization reasons we only plot 10\% of all data points. One can see a clear asymmetry toward fainter magnitudes showing that artificial stars are more often recovered with magnitudes brighter than their input caused by significant blending in our images.}
	\label{fig:photometric_uncertainty}
\end{figure*}

The photometric uncertainties provided by DOLPHOT (Paper I) are the uncertainties calculated from the signal-to-noise ratio (S/N) of the measurements. These uncertainties are a lower limit to the real uncertainties, which are also influenced by the local environment of a source, such as crowding and extinction. The artificial star tests give us the opportunity to properly determine the real photometric uncertainties, since we know the input magnitude and the measured output magnitude of the recovered stars. The photometric uncertainty can be estimated by using the scatter of the measured photometry of the artificial stars. In Fig.~\ref{fig:photometric_uncertainty} we show the difference between the input and measured photometry for each of the wide-band filters. Negative values are due to the measured magnitude being smaller than the input one, implying that the star appears brighter.

The absolute magnitude difference  $\Delta m_i$  for each recovered artificial star $i$ is calculated as:
\begin{equation}
\label{eq:absolute spread}
\Delta m_i=\Vert m_i^{\rm out}-m_i^{\rm in}\Vert.
\end{equation}
 
After fitting an exponential to the scatter values as a function of magnitude, we can assign the real photometric uncertainty for each star in each of the four wide-band filters and compare it with the uncertainty obtained in Paper~I from the observations with DOLPHOT at the magnitude below which 10\%, 50\%, and 90\% of all sources are found. We list both uncertainties for each filter in Tab.~\ref{tab:phot_uncertainty}. As one can see, the uncertainties derived from the S/N are consistently smaller, especially for faint magnitudes.

\floattable
\begin{deluxetable}{cccccccccc}
	\tablecaption{The photometric uncertainty \label{tab:phot_uncertainty}}
	\tablehead{
			\multicolumn{1}{c}{Filter} &\multicolumn{3}{c}{mag-limit [mag]} & \multicolumn{3}{c}{$\sigma_m$~[mag]} & \multicolumn{3}{c}{$\Delta m$~[mag]}\\
			\multicolumn{1}{c}{ } &\multicolumn{1}{c}{10\%} & \multicolumn{1}{c}{50\%} & \multicolumn{1}{c}{90\%}  &\multicolumn{1}{c}{10\%} & \multicolumn{1}{c}{50\%} & \multicolumn{1}{c}{90\%}  &\multicolumn{1}{c}{10\%} & \multicolumn{1}{c}{50\%} & \multicolumn{1}{c}{90\%}
		}
	\startdata
	$F555W$ & 19.3  &  24.5 	& 26.7 & 0.002  &  0.037   & 0.160 & 0.003	& 0.052  &   0.216	\\
	$F814W$ & 19.4  &  22.7 	& 25.3 & 0.002  &  0.011   & 0.093 & 0.009	& 0.048  &   0.197	\\
	$F125W$ & 17.2  &  19.9 	& 22.7 & 0.001  &  0.005   & 0.035 & 0.016	& 0.060  &   0.240	\\
	$F160W$ & 16.6  &  19.8 	& 22.6 & 0.001  &  0.006   & 0.054 & 0.017	& 0.087  &   0.330	\\
	\enddata
	\tablecomments{The uncertainties for each of the broad-band filters at the magnitude below which 10\%, 50\%, and 90\% of all sources are found. Column~2 gives the magnitude while columns~3--5 and 6--8 list the magnitude uncertainties determined with DOLPHOT (see Paper~I) and the uncertainties determined via the artificial star test, respectively.}
\end{deluxetable}

\section{The spatial distribution of the stellar population}
\label{sec:dist}

In Paper~I, we showed that the stellar population of RCW~49 mainly consists of PMS stars and massive OB main-sequence stars. We also showed that the Wd2 cluster contains two subclusters, the main cluster (MC) and the northern clump (NC). After the preliminary estimate of the clump sizes and locations in Paper~I, we now perform a more sophisticated analysis of the spatial distribution also taking into account completeness effects. Due to its location in the Galactic disk, we had to separate the Wd2 cluster population from the foreground field population. This was done by a simple linear cut in color-magnitude space using the $F814W-F160W$ vs. $F814W$ CMD (see Fig.~16 and Fig.~17 in Paper~I).

\subsection{The spatial distribution of the stellar population}
\label{sec:spatial_distribution}

\citet{Vargas_Alvarez_13}, \citet{Hur_15}, and \citet{Zeidler_15} pointed out that Wd2 shows a smaller clump northwards of the MC. \citet{Hur_15} named this concentration of stars 'northern clump'. After a preliminary analysis of the spatial extent and the stellar content of the two clumps without completeness correction, we found that the MC contains roughly twice as many stars as the NC (Paper~I). The respective CMDs do not show any pronounced difference in their stellar populations or their age (0.5--2~Myr, Paper~II).

In order to determine the surface density distribution of the stellar population of Wd2, we corrected our photometric catalog for completeness as described in Sect.~\ref{sec:completeness}. Although the level of completeness is very high across the cluster area (see Sect.~\ref{sec:completeness_maps}), in the cluster center, very close to the O and B stars, the detection rate drops rapidly in all filters, caused by crowding.

In order to correct properly for incompleteness, brightness limits are needed in the two filters used for the selection of cluster members. These are $F814W=21.8$~mag and $F160W=17.8$~mag and correspond to $M \sim 0.5M_\odot$. Thus stars fainter than these values are not considered. These limits were chosen so that in the center of the clumps a completeness level of 50\% is reached.

We then calculated the surface density of the remaining cluster members, obtaining the local, completeness-corrected stellar surface density around each star (within a radius of 10.8~arcsec) of all point sources brighter than $F814W=21.8$~mag and $F160W=17.8$~mag. These rather irregular positions of the point sources were then triangulated to a regular grid of $392 \times 403$ bins (binsize: 1.08~arcsec), using the \texttt{Triangulate} and \texttt{TriGrid} routines in IDL\footnote{Additional data analyses were done using IDL version 8.5 (Exelis Visual Information Solutions, Boulder, Colorado).}.

To determine the surface density profile of Wd2, as well as the shapes and spatial extents of the two clumps, we fitted a combination of two 2D-Gaussians. The peak coordinates (J2000) for the MC are R.A.=$10^{\rm h}24^{\rm m}01^{\rm s}.65$ and decl.=$-57^\circ45'33.4''$ with a peak density of 1863~stars~arcmin$^{-2}$. For the NC we obtained R.A.=$10^{\rm h}24^{\rm m}02^{\rm s}.16$ and decl.=$-57^\circ44'39.3''$ with a peak density of 937~stars~arcmin$^{-2}$. This leads to a projected distance between the peaks of the two clumps of $d=54.25$~arcsec \citep[1.09~pc at the distance of 4.16~kpc,][]{Vargas_Alvarez_13,Zeidler_15}.

To analyze and compare the properties of the two clumps, we defined the $1\sigma$ width of the fitted Gaussian distributions as the size of the two clumps. An elliptical shape distribution for the MC is not a very accurate representation and for this reason we used the contour line at the $1\sigma$ width of the Gaussian fit. The NC partly overlaps with the wings of the MC distribution. Therefore, we could not use a closed contour line for the $1\sigma$ density. Instead we approximated it with an ellipse, which is the result of the 2D Gaussian fit. The major and minor axis represent the $1\sigma$ width of the Gaussian distribution. Within these boundaries, the MC and NC cover an area of 0.31~arcmin$^{2}$ (0.44~pc$^2$) and 0.27~arcmin$^{2}$ (0.38~pc$^2$), respectively.

To separate the Wd2 cluster from the surrounding \ion{H}{2} region of RCW~49, we defined its $2\sigma$ density contour as the cluster boundary. Its projected area is 2.57~arcmin$^{2}$ (3.74~pc$^2$). The sizes of our clumps ($r_{\rm MC}=0.37$~pc and $r_{\rm NC}=0.35$~pc) are comparable with the core radii that \citet{Kuhn_14} derived in their analysis of the subclustering of 17 regions (unrelated to Wd2) in the Massive Young Star-Forming Complex Study in Infrared and X-ray \citep[MYSTiX, see][]{Feigelson_13} project. The sizes of the cores in their study range from 0.01~pc to $>2$~pc, where 68\% of their core radii are between 0.06 and 0.45~pc.

The MC includes 378 observed PMS members (490 after completeness correction) leading to a completeness-corrected surface density of $\Sigma_{MC}=1114$~stars~pc$^{-2}$, while the NC comprises 157 observed PMS members (211 after completeness correction) resulting in a completeness-corrected surface density of $\Sigma_{NC}=555$~stars~pc$^{-2}$. The surface density of the Wd2 cluster as a whole, including MC and NC, is $\Sigma_{Wd2}=264$~stars~pc$^{-2}$. We note here again that all surface densities are based on a lower mass cutoff at $0.5M_\odot$.

\subsection{The uniformly distributed low-mass stellar population}
\label{sec:low-mass_pop}

We analyzed the observed and completeness-corrected spatial distributions of the stellar population in RCW~49 in different brightness bins. Outside the two clumps we can analyze stars with masses lower than $0.5M_\odot$. The spatial distribution of the lowest-mass stars ($<0.15~\rm{M}_\odot$ down to the detection limit of $\sim 0.09~\rm{M}_\odot$) outside the two clumps appears uniform throughout the RCW~49 region. In total, 1293 low-mass cluster members ($<0.15~\rm{M}_\odot$) are detected. This gives a total, completeness-corrected number of 1494 low-mass stars, down to the detection limit $\sim 0.08~$M$_\odot$,  defining a stellar surface density of $\Sigma=105~\rm{stars~pc}^{-2}$, not including the regions of the NC and the MC (see Sect.~\ref{fig:completeness}).

This uniformly distributed low-mass stellar population extends to the ridges of the surrounding gas and dust cloud. Additionally, we found in Paper~II a small age gradient of $\sim 0.15$~Myr decreasing from the Wd2 center toward the gaseous ridge, which suggests that the OB star population of Wd2 might be triggering star formation in the surrounding gas cloud.

This phenomenon was already observed by \citet{Whitney_04} who identified $\sim 300$ candidate young stellar objects (cYSOs) in the giant \ion{H}{2} region of RCW~49 using data obtained with the Infrared Array Camera \citep[IRAC,][]{Fazio_04} on board the \textit{Spitzer} Space Telescope \citep{Werner_04} for the Galactic Legacy Infrared Mid-Plane Survey Extraordinaire (GLIMPSE). The observations were carried out in the IRAC bands 1 to 4 with central wavelengths of 3.6~$\mu$m, 4.5~$\mu$m, 5.8~$\mu$m, and 8.0~$\mu$m, respectively. Further details on this program are provided in \citet{Benjamin_03}. \citet{Whitney_04} concluded that the central cluster had induced a second generation of star formation through feedback, similar to what was observed in NGC~602 in the Small Magellanic Cloud by \citet{Carlson_07}. However, because of the limited spatial resolution of the $Spitzer$ Space Telescope, the lower mass limit of the observations of \citet{Whitney_04} is $2.5M_\odot$. We calculated that, using our data in the central area, this limit translates into missing the vast majority ($\sim 95\%$) of PMS stars.

\subsection{The red tail of the PMS population}
\label{sec:red_tail}

\begin{figure*}[htb]
	\resizebox{\hsize}{!}{\includegraphics{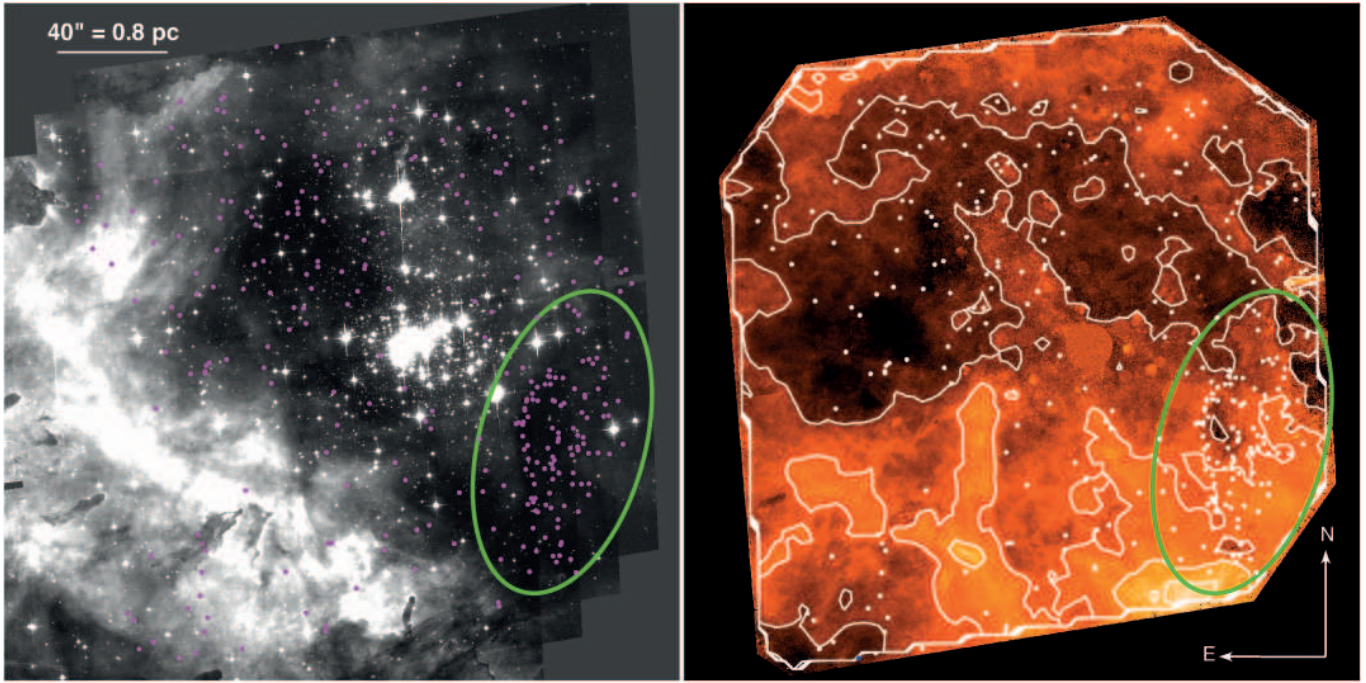}}
	\caption{The left panel shows the $F814W$ image (corresponding to the $I$-band) of the Wd2 region. The right panel shows the $E(B-V)_g$ color-excess map created in Paper~I. Over-plotted are (in magenta (left) or white(right)) the objects of the red tail. The green ellipse marks the region of overdensity.}
	\label{fig:F814W+redmap}
\end{figure*}

The dereddened $F814W_0$ vs. $(F814W-F160W)_0$ CMD shows 304 faint, very red objects with $F814W_0 > 18.0$~mag and $(F814W-F160W)_0 > 3.2$~mag. In Fig.~\ref{fig:F814W+redmap}, we overplotted these selected objects as magenta dots on the $F814W$ image of our observations and our 2D high-resolution reddening map (see Paper I). These red stars are concentrated in the south-west corner of the FOV. Since all these objects were individually dereddened using the gas extinction map, the distribution should be uniform throughout the field (assuming that the gas extinction map provides a fair representation of also the stellar extinction). The right panel of Fig.~\ref{fig:F814W+redmap} shows the reddening map including nine contour levels with linear steps between $0 \le E(B-V)_g \le 2.89$~mag. No obvious spatial correlation between the red objects (marked in both frames within the green ellipse) and the distribution of the gas and dust is visible. On the other hand, the gas excess map only represents the nebular extinction (through the H$\alpha$ and Pa$\beta$ line emission). Within the ellipse, the median $E(B-V)_g$ is with 2.1~mag, 0.23~mag higher than the global median of the gas extinction map (see Paper I).

\subsubsection{Moving toward the mid-infrared}

\begin{figure*}[htb]
	\resizebox{\hsize}{!}{\includegraphics{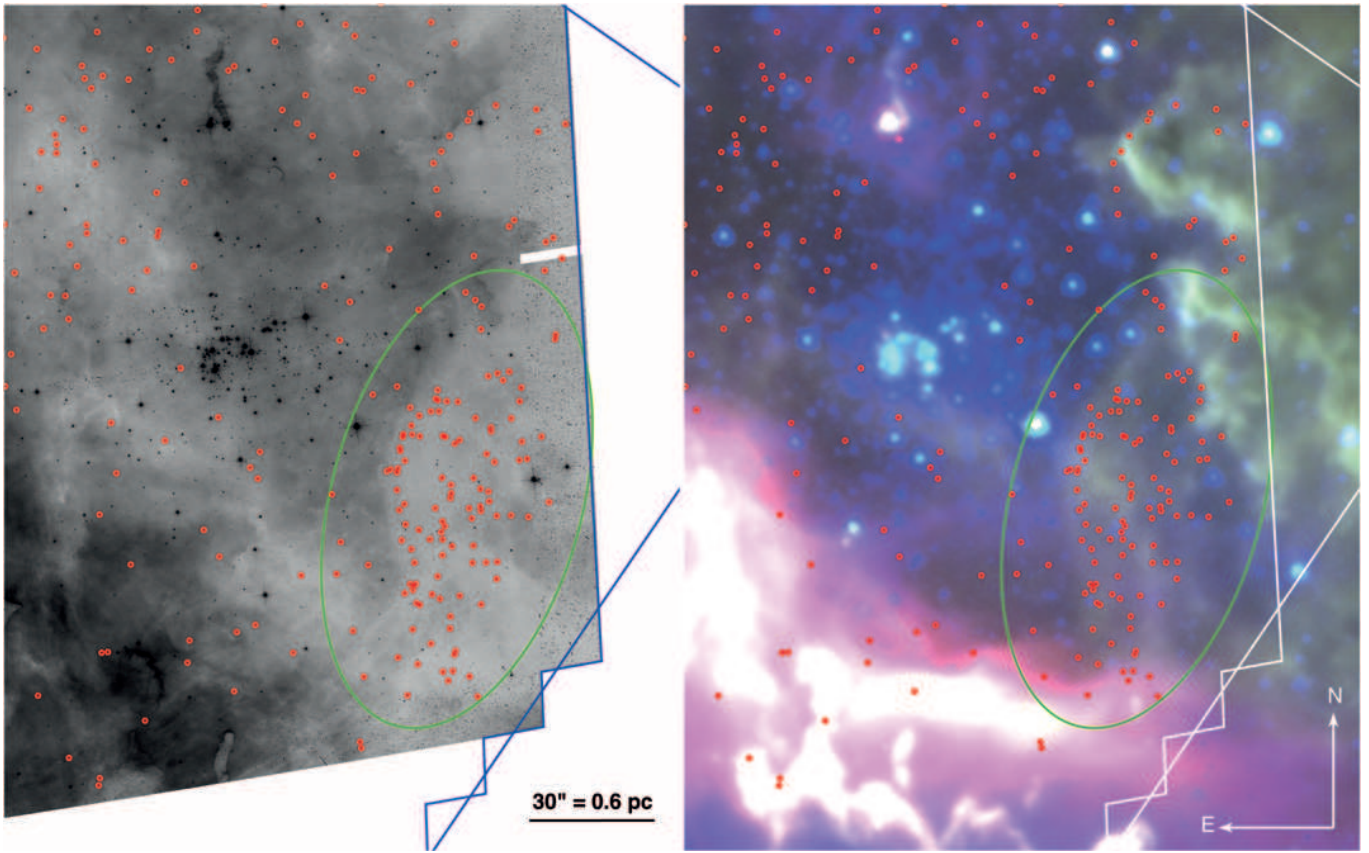}}
	\caption{In the left panel we show the inverted $F658N$ image from our observations, mapping the H$\alpha$ emission. Dark regions are bright in H$\alpha$. The right panel shows the three color-composite image from the \textit{Spitzer} GLIMPSE survey \citep{Benjamin_03}. The [8.0], [5.8], and [4.5]-bands are shown in red, green, and blue, respectively. The blue (left) and white (right) line trace the FOV of our HST observations. The objects in the red tail in the CMD are overplotted in red. The green ellipse marks the region of the red tail overdensity (same as in Fig.~\ref{fig:F814W+redmap}). The objects in the red tail overdensity are located in a region of negligible H$\alpha$ emission, while there is PAH emission visible in green in the \textit{Spitzer} image.}
	\label{fig:Halpha+Spitzer}
\end{figure*}

The filter with the longest wavelength available in our HST survey is the $F160W$ filter, corresponding to the $H$-band in the NIR. To detect molecular gas and dust clouds the use of filters with wavelengths in the mid-infrared (MIR) and sub-millimeter becomes necessary. In Fig.~\ref{fig:Halpha+Spitzer} we show the zoomed-in region in which the red tail stars (over-plotted as red points) are located. The left panel shows the inverted $F658N$ image, sampling the H$\alpha$ emission. In the right panel we show a three-color composite image using the IRAC 2 ($4.5~\mu$m), 3 ($5.8~\mu$m), and 4 ($8.0~\mu$m) \textit{Spitzer} observations for blue, green, and red, respectively, based on data from the GLIMPSE survey \citep{Benjamin_03}. The ellipse marks the region of the overdensity and is identical to the one in Fig.~\ref{fig:F814W+redmap}.

From the examination of the $Spitzer$ color-composite image, we see a molecular cloud at the same location. The [3.6], [5.8], and [8.0] IRAC bands show different emission features of polycyclic aromatic hydrocarbons (PAHs). The [3.6], [5.8], and [8.0] bands contain the $3.3~\mu$m, $6.2~\mu$m, and $7.7~\mu$m C--H stretching modes, respectively. The [8.0] band additionally contains the $8.6\mu$m C--H in-plane bending mode \citep{Draine_03}, while the IRAC [4.5] band contains the bright Br$\alpha$ hydrogen line (centered at $4.05~\mu$m). \citet{Churchwell_04} predicted that the Br$\alpha$ emission contains $\sim20\%$ of the non-stellar flux observed in \object{RCW~49}. Looking at the colors of the right panel in Fig.~\ref{fig:Halpha+Spitzer}, the clouds containing the overdensity of the red tail objects appear to be dominated in the [5.8] IRAC-3 band by PAH emission produced by molecules with C and O atoms in their structure. In areas where we can see the Br$\alpha$ emission (blue), we also can see a higher flux in the H$\alpha$ image (left frame).

\subsubsection{The $^{12}$CO and $^{13}$CO transition lines}

The properties of the molecular gas associated with Wd2 have been analyzed in several studies. The major difficulty is that Wd2 is located close to the tangent point of the Carina Arm as seen from us \citep{[i.e.][]Grabelsky_87}, meaning that for gas clouds with a distance to the Sun smaller and larger than Wd2 the distance determination is degenerate. The studies of the CO emission spectrum in combination with the 21~cm absorption spectrum by \citet{Dame_07} suggest that Wd2 must be located in between two molecular clouds. Wd2 was studied in more detail by \citet{Furukawa_09,Furukawa_14} and \citet{Ohama_10} using  the NANTEN~2 4m submillimeter/millimeter telescope \citep[from the NANTEN CO Galactic Plane Survey,][]{Mizuno_04} in combination with \textit{Spitzer} data. They detected three different CO clouds with 3 different velocities (16, 4, and $-4$~km~s$^{-1}$, relative to the local standard of rest (LSR), see Fig.~1 in \cite{Ohama_10} and Fig.~1 in \cite{Furukawa_09}). The cloud with a peak velocity of $-4$~km~s$^{-1}$ spatially coincides very well with the location of the molecular cloud where the red tail overdensity is located (see bottom right of Fig.~1 in \citet{Furukawa_09} and bottom panels in Fig.~1 of \citet{Ohama_10}). Using the rotation curve of \citet{Brand_93}, \cite{Furukawa_09} obtained kinematic distances of 6.5, 5.2, and 4.0~kpc for the 16, 4, and $-4$~km~s$^{-1}$ clouds, respectively. \citet{Furukawa_09,Furukawa_14} and \citet{Ohama_10} also suggested that the creation of the RCW~49 region, with the Wd2 cluster as its central ionizing cluster, was induced by a cloud-cloud collision of two of these observed molecular clouds. This collision may have caused the formation of the spatially uniformly distributed low-mass PMS stars that we see in our HST data. Such a mechanism can induce star formation throughout the molecular cloud. 

As we can only see an additional reddening in the region where the $-4$~km~s$^{-1}$ molecular cloud is located, we conclude that this cloud lies in front of Wd2. This indicates that the other clouds are located in the background of Wd2, which would confirm our determined distance of 4.16~kpc (Paper~I). Thus any extinction caused by these clouds would only affect objects located behind the cluster as seen from us and hence be undetectable in our data. The cloud kinematics of \citet{Dame_07} and \citet{Furukawa_09} suggest that Wd2 is located behind the 4, and $-4$~km~s$^{-1}$ clouds, which is in contradiction to our distance estimate. On the other hand, if we underestimated the extinction toward Wd2, our distance modulus is then overestimated leading to an even closer distance. We suggest that Wd2 is located between the 4 and $-4$~km~s$^{-1}$ clouds confirming our distance of 4.16~kpc as determined in Paper I.

The stars of the red tail are probably RCW~49 PMS stars that are still not corrected completely for extinction and, therefore, appear too red. Because the molecular gas is not located directly along the line of sight to Wd2, we can also conclude that for the remaining part of RCW~49 in our observed FOV, our method for the correction for extinction is not significantly contaminated by molecular extinction.

\section{The present-day mass function}
\label{sec:mass_function}

The only published PDMF of Wd2 was determined by \citet{Ascenso_07} with ground-based photometry in the NIR ($J$, $H$, $K_s$), who derived a slope of $\Gamma=-1.20 \pm 0.16$. The slope is based on the mass range between 0.8--25$~M_\odot$, representing their completeness limits, for an assumed distance of 2.8~kpc. Using our data we are able to refine the PDMF of Wd2 with a much higher accuracy.

\subsection{The stellar masses}
\label{sec:stellar_masses}

\begin{figure*}[htb]
	\resizebox{\hsize}{!}{\includegraphics{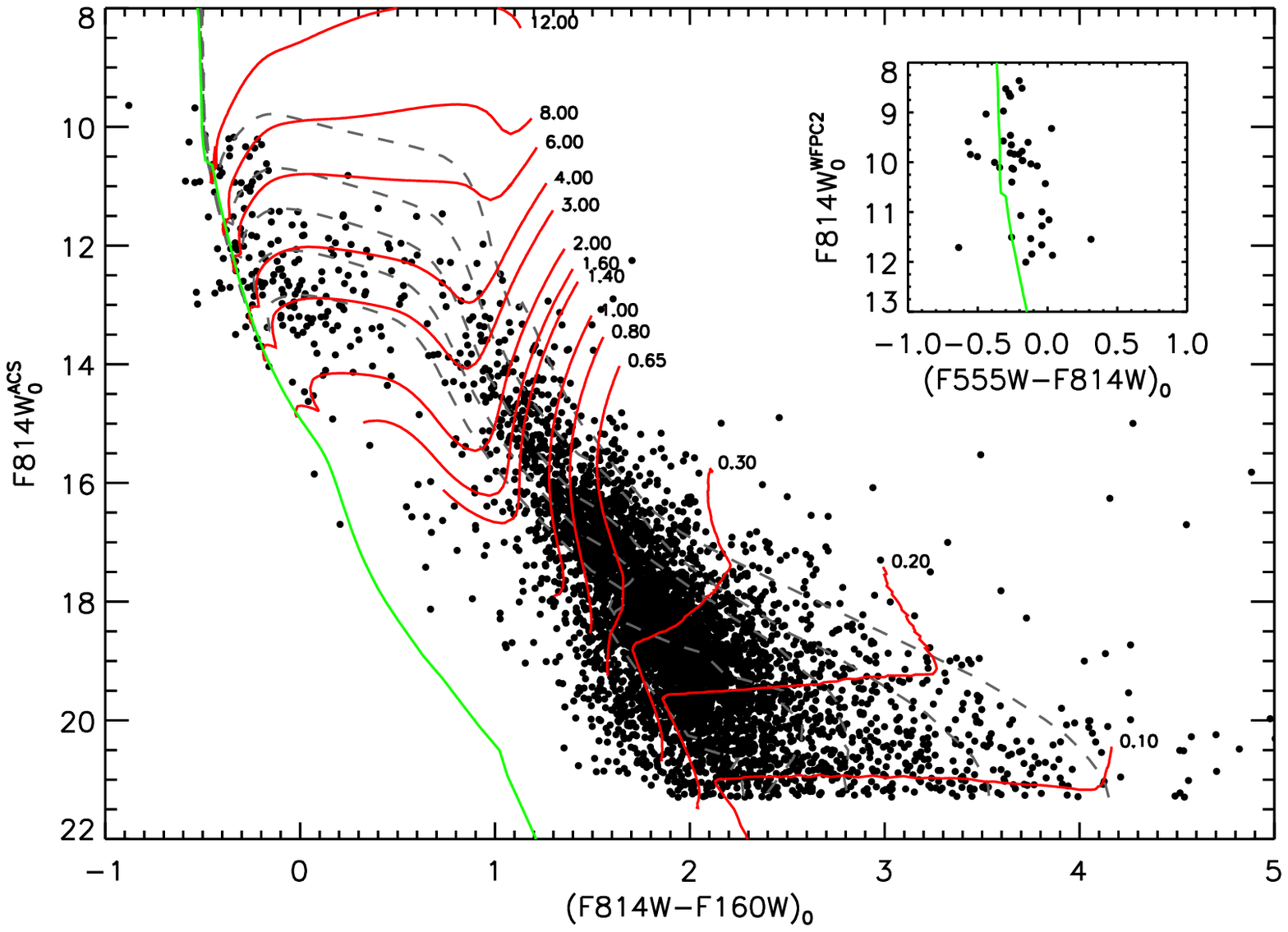}}
	\caption{The $F814W_0$ vs. $(F814W-F160W)_0$ CMD. The evolutionary tracks of \citet{Bressan_12} and \citet{Bressan_13} are traced in red and are labeled with their corresponding masses. The gray dashed lines are the PMS isochrones for 0.1, 0.25, 0.5, 1.0, and 2.0~Myr. Additionally we show in the right top corner the $F814W_0$ vs. $(F555W-F814W)_0$ CMD of the 43 stars added using the WFPC2 photometry from \citet{Vargas_Alvarez_13}, which includes stars that are saturated in our data. The green line represents the ZAMS, inferred from the same models.}
	\label{fig:CMD_evo_tracks}
\end{figure*}

To determine the PDMF of Wd2 we need to carefully estimate the masses from the stellar luminosities. For MS stars belonging to single stellar populations this can be done via their locus relative to an isochrone since for MS stars their luminosity is closely correlated with stellar mass. Uncertainties are added by (unknown) stellar rotation and possible binarity and multiplicity. In Paper~I and II we showed that the RCW~49 region mainly consists of PMS stars with an age distribution between 0.1--2.0~Myr. However, the locus of the PMS stars in a CMD is not only a function of age, but is also affected by ongoing accretion, presence and viewing angle of protostellar disks, emission lines, etc. To remove the age-dependence of the PMS stars we use evolutionary tracks instead of isochrones. PMS evolutionary tracks have the advantage that they provide us with all the information about the stellar masses relative to their locus in the CMD but are independent of an age-binning, since they follow the evolution of a star with a specific mass on the Hayashi track \citep{Hayashi_61}.

In order to estimate the stellar masses we determined the locus of each star relative to the evolutionary tracks of \citet{Bressan_12} and \citet{Bressan_13} in combination with the PARSEC 1.2S isochrones. We used mass bins that have a similar distance in $\log(m)$ (see Fig.~\ref{fig:CMD_evo_tracks}). We added 43 MS stars from the photometric catalog of \citet{Vargas_Alvarez_13} obtained with WFPC2. These are bright MS stars saturated in our observations. Typical uncertainties for stars $<15$~mag are $<0.01$~mag \citep{Vargas_Alvarez_13}, comparable to the uncertainties of our photometric catalog and so the combination of the two photometric catalogs does not introduce additional uncertainties. We performed a visual inspection of all stars that are saturated in our data and confirmed that they are covered by the photometric catalog of \citet{Vargas_Alvarez_13}. All stellar masses for stars bluer than the
MS and redder by at most $\le 0.1$~mag than the MS, as well as brighter than $F814W_0 = 14$~mag were determined via their locus relative to the zero-age main sequence (ZAMS) instead of evolutionary tracks, because close to the ZAMS the mass determination via evolutionary tracks is degenerate. In Fig.~\ref{fig:CMD_evo_tracks} we show the cluster members overplotted on the evolutionary tracks (in red) and the ZAMS (in green). In the top right corner we show the CMD for the 43 MS stars added from the WFPC2 photometry of \citet{Vargas_Alvarez_13} as an inset. To determine the masses for the stars with WFPC2 photometry, we used the ZAMS in the WFPC2 photometric system to avoid introducing uncertainties due to different photometric systems.

\begin{figure*}[htb]
	\resizebox{\hsize}{!}{\includegraphics{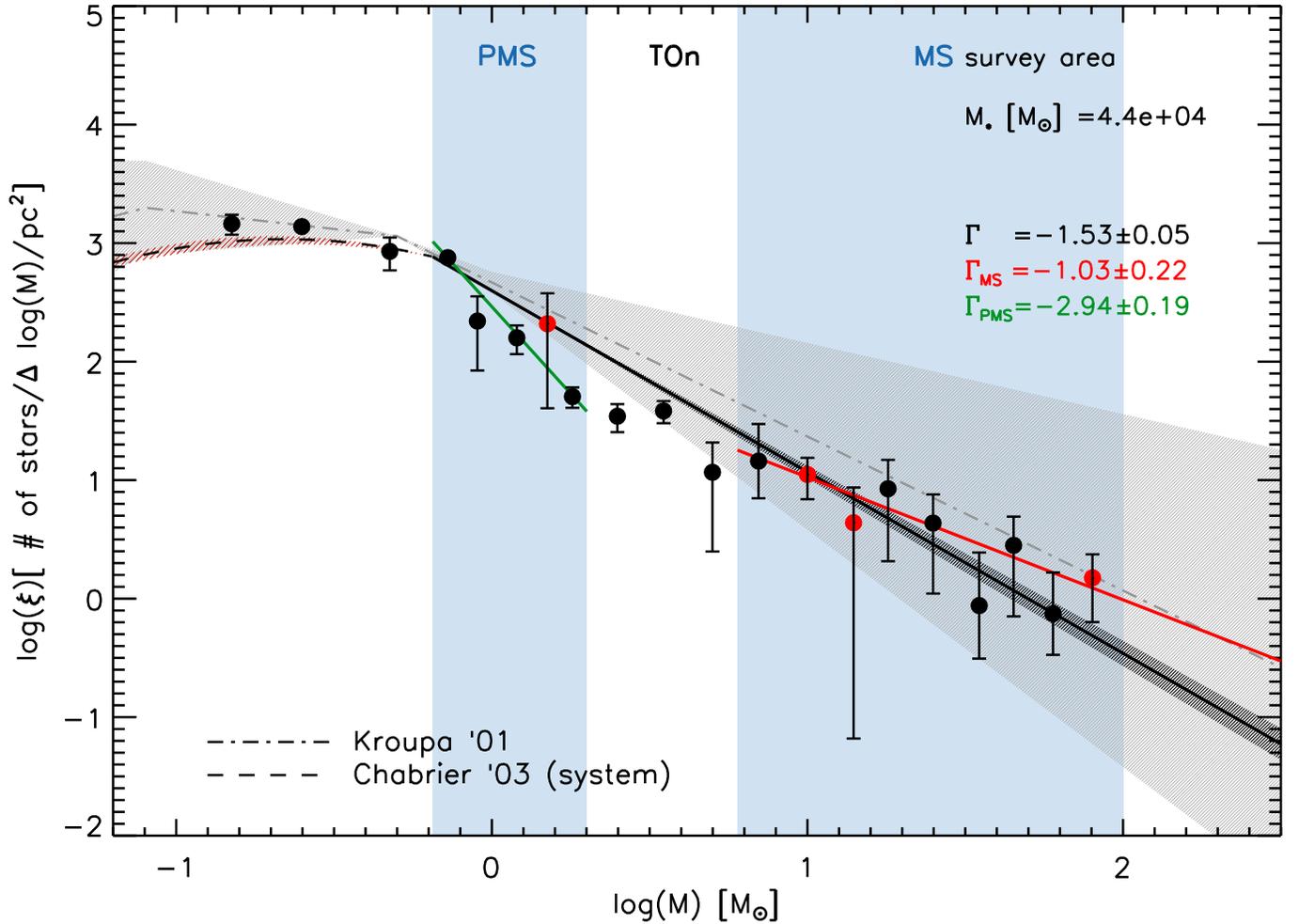}}
	\caption{The PDMF of the full survey area of our HST data. The black filled circles represent the completeness-corrected, normalized number of stars per logarithmic mass bin. The red filled circles stand for those bins for which at least one star's mass is estimated via orbits of the spectroscopic binary components from the literature. The two blue areas are the regions where the PDMF fit is performed. The black solid line represents the error-weighted fit of the overall PDMF with the uncertainty marked as shaded area. The dash-dotted line is the \citet{Kroupa_01} IMF with the associated uncertainty as gray area. The dashed line is the \citet{Chabrier_03a} IMF with the corresponding uncertainty as reddish area \citep{Chabrier_03b} for the low-mass stars. The green and red solid lines indicate the PMS and MS PDMF, respectively.}
	\label{fig:MF_FOV}
\end{figure*}

In order to compute the PDMF of Wd2 we follow the technique presented by \cite{Tarrab_82} and \citet{Massey_95}, also adopted by \citet{Sabbi_08} and \citet{Cignoni_09}, where stars lying between two evolutionary tracks of different initial mass are assigned to a mass bin. We used the mass bins described above, which were chosen to establish a regular grid in $\log(m)$. The number of stars per mass bin was also corrected for completeness (see Sect.~\ref{sec:completeness}). We emphasize here that we corrected each star individually with its specific completeness fraction, depending on its magnitude, color, and position. Some stars (three for the survey area) are saturated in all but one filter ($F555W$), in both our images and the WFPC2 photometry of \citet{Vargas_Alvarez_13}. Their spectral types \citep[determined in ][]{Rauw_05,Rauw_07,Rauw_11,Vargas_Alvarez_13} confirm their MS membership and as a result we were able to determine their masses by comparing only the $F555W$ photometry to the ZAMS.

For three binary systems the orbits of the stellar components are known and the mass determination for the members of these binaries is accurate, so we used these masses. Analyzing in total 23 spectral observations, \citet{Rauw_04} found that WR20a is a double-lined binary with a preferred orbital period of $3.675 \pm 0.030$\,days and with component minimum masses of $68.8 \pm 3.8\,\rm M_\odot$ and $70.7 \pm 4.0\,\rm M_\odot$. They suggest a spectral type of WN6ha for both components. \citet{Bonanos_04} discovered it was also an eclipsing binary and by measuring its inclination they were able to refine its component masses to even higher values of $ 83.0 \pm 5.0\,\rm M_\odot$ and $82.0 \pm 5.0\,\rm M_\odot$. For MSP~96 \citep[see ][]{Moffat_91}, \citet{Rauw_11} determined orbital solutions using VLT/FLAMES spectra. They determined an orbital period of $1.0728\pm0.0006$\,days. They suggest minimum masses of $M_p \sin^3i=11.1\pm 3.0\,\rm M_\odot$ and $M_s \sin^3i=13.8\pm 6.6\,\rm M_\odot$ for the primary and secondary component, respectively. Their $BV$ photometry supports absolute masses of $M_p=12.0\pm 3.3\,\rm M_\odot$ and $M_s=15.0\pm 7.2\,\rm M_\odot$ and a spectral type of B1V is suggested for both components. For the binary MSP~44, \citet{Rauw_11} determined an orbital period of $5.176\pm0.029$\,days. From the very low mass ratio, they inferred a B1 primary with a mass of 10--12\,M$_\odot$ and a PMS companion with a mass of 1.4--1.6\,M$_\odot$, most likely being an F-type star.

\subsection{The slope of present-day mass function}
\label{sec:mass_funct_slope}

The MF is defined as the fractional number of stars per mass interval per unit area. A common parametrization is the one used by \citet{Scalo_86}, in the form of $\Gamma = d\log\xi\left(\log(m)\right)/d\log(m)$, where $m$ is the stellar mass and  $\xi\left(\log(m)\right)$ the number of stars per logarithmic mass interval. For the Solar neighborhood the slope of the MF is $\Gamma=-1.35$, as derived by \citet{Salpeter_55}. In Fig.~\ref{fig:MF_FOV} we present the MF for the survey area (thick black dots). Mass bins that contain at least one star whose mass is inferred by orbital parameters from spectroscopy are indicated by red filled circles. There are six objects that were identified spectroscopically as binaries or binary candidates but the individual components are unresolved and their masses unknown. Two objects are confirmed binaries, namely MSP~223 \citep[which is also an eclipsing canditate, see ][]{Rauw_07} and MSP~188 \citep{Vargas_Alvarez_13}. Four objects are binary candidates: MSP~18 and MSP~167a, classified by \citet{Rauw_11}, and MSP~175 and star \#714\footnote{This star is not included in \citet{Moffat_91} and so lacks a MSP identifier. Its number is the identifier of \citet{Vargas_Alvarez_13}.}, classified by \citet{Vargas_Alvarez_13}. If we treat the photometry of the unresolved binaries in the same way as the remaining single stars, the slope of the mass function changes by 2.2\% ($\Delta \Gamma = 0.0332$). This shows that excluding these binaries does not significantly affect the PDMF.

To determine the present-day MF (PDMF) we used two regions to fit the slope: the PMS above $0.65~M_\odot$ and below $2.0~M_\odot$, where the completeness is at least 50\% everywhere, and the MS for all stars between $6.0~M_\odot$ and $50~M_\odot$. Between $2.0~M_\odot$ and $6.0~M_\odot$ most stars are in the turn-on (TOn) region\footnote{The turn-on region is the transition region in the CMD, where stars evolve from the PMS onto the ZAMS.} (see Fig.~\ref{fig:CMD_evo_tracks}), for which the uncertainties of the stellar evolution models are high, and therefore we excluded this region from the fit. We used an error-weighted linear fit. For the complete survey area we obtain a PDMF slope of $\Gamma=-1.53 \pm 0.05$. The slope of the MS is $\Gamma_{\rm{MS}}=-1.03\pm 0.22$. The slope based on the PMS is $\Gamma_{\rm{PMS}}=-2.94 \pm 0.19$, thus much steeper (see Fig.~\ref{fig:MF_FOV}). The much shallower slope of the PDMF in the MS mass regime as compared to the PDMF computed for the PMS and all stars suggests the existence of mass segregation but uncertainties in the inferred PMS masses may also contribute (see Sect.~\ref{sec:uncertainties}). We will discuss this in more detail in Sect.~\ref{sec:mass_segregation}. For the Wd2 cluster ($2\sigma$ boundary, see Sect.~\ref{sec:spatial_distribution}) we obtain a PDMF slope of $\Gamma=-1.46 \pm 0.06$. The slope of the MS is shallower with $\Gamma_{\rm{MS}}=-1.01\pm 0.22$. The slope based on the PMS is $\Gamma_{\rm{PMS}}=-2.89 \pm 0.24$.

\subsection{Uncertainties of the PDMF}
\label{sec:uncertainties}
To estimate the uncertainties influencing the determination of the PDMF we list the individual sources of uncertainties:

\begin{itemize}
\item The photometric errors in each filter, determined via the artificial star tests as described in Sect.~\ref{sec:true_phot_uncert}. They affect the luminosity as well as the color of each star and so the stellar mass.
\item The uncertainty on the stellar color-excess used to deredden the photometry (see Paper~I).
\item The uncertainty on the distance to Wd2 of $\delta d=0.33$~kpc \citep{Vargas_Alvarez_13,Zeidler_15} resulting in a distance modulus uncertainty of 0.175~mag.
\item The statistical uncertainty on the number of stars per mass bin estimated with the Poisson statistics.
\item Stellar rotation (although this effect is less severe for low-mass stars, \citet{Bastian_09})
\item Unresolved binary systems
\item High uncertainties of the PMS star masses due to variable accretion, different circumstellar disk sizes, and different inclination angles, etc.
\end{itemize}

To estimate the uncertainties per mass bin we varied the stellar luminosities according to the uncertainties and estimated the change in the number of stars per mass bin. The photometric uncertainties are more dominant in the low-mass regime, while the statistical sampling (Poisson error) dominates the uncertainties at the high-mass end, where only a few stars are left. For the added binary system components with spectroscopically inferred masses, we used the published mass uncertainties \citep[][and references therein]{Vargas_Alvarez_13}. The total uncertainties are shown by the black error bars in Fig.~\ref{fig:MF_FOV}. The uncertainties propagated to the slope of the PDMF are represented by the gray-shaded area around the PDMF (black solid line).

PMS stellar evolutionary models carry large uncertainties originating from the inaccuracies of the model atmospheres due to necessary simplifications of the physical processes \citep[e.g.,][]{Preibisch_12a}. Thus mass and age determinations, especially in the low-mass regime for young ages, can come with large uncertainties. Throughout this paper we used the PARSEC 1.2S isochrones and evolutionary tracks \citep{Bressan_12}. This code has the advantage of computing isochrones for the different filter sets of all of the major telescopes, including HST. The MESA Isochrones and Stellar Tracks \citep[MIST,][]{Dotter_16,Choi_16} provide similar information. Down to masses of $0.5~{\rm M}_\odot$ the MIST isochrones and evolutionary tracks agree very well with the PARSEC 1.2S ones. Our lower mass limit to determine the PDMF and total masses is $0.65~{\rm M}_\odot$, caused by incompleteness effects, so these deviations at the low-mass end are minor for our purpose. Also the PARSEC 1.2S isochrones follow better our observed low-mass data points, which prompted us to adopt them. Nevertheless, we used the MIST evolutionary track to estimate the model uncertainties. We computed the slope of the mass function and the total mass of the survey area (see next Sect.~\ref{sec:tot_mass}) using the MIST model. The slope of the PDMF deviates by $\sim 12\%$, while the total mass deviates by $\sim 17\%$. We introduce these as the model uncertainties.

Different studies showed that the majority of the massive stars are binaries \citep[e.g., ][]{Sana_12,Sana_13}. For Wd2 we only have an incomplete census of the binary systems that can introduce additional uncertainties to the PDMF. To address the unknown binary fraction we assume the extreme case scenario that all stars $\ge 8\,\rm M_\odot$ are equal mass binaries. This leads to a PDMF slope that is steeper by $\sim 0.02$, suggesting that treating these stars as single stars does not introduce a large uncertainty. Nevertheless, we added the estimated uncertainty due to possible binarity to the individual uncertainties of the mass bins, meaning that we determined the difference in number of stars in each bin with mass $\ge 8\,\rm M_\odot$ between the case where all stars are single and the case where all stars are equal-mass binaries. We have to note here that this approach shows the extreme case and, therefore, the introduced uncertainties represent an upper limit.

\subsection{The total stellar mass of Wd2}
\label{sec:tot_mass}
To determine the stellar mass of Wd2 and of the RCW~49 region we separated the stellar distribution into two mass regimes ($M \ge m_{50}=0.65~M_\odot$ and $M < m_{50}=0.65~M_\odot$), where $m_{50}$ is the stellar mass at the 50\% completeness limit for a 1~Myr old star. For the mass range above the 50\% completeness limit the total mass is calculated by simply summing up the completeness-corrected numbers multiplied by the masses $m_{\rm{tot}}=\sum m_i \cdot n_i$. To estimate the integrated stellar mass below the 50\% completeness limit we used the \citet{Chabrier_03a} IMF (for unresolved binaries, see Fig.~\ref{fig:MF_FOV}) in the following form:

\begin{equation}
\label{eq:chabrier_IMF_log}
\xi(\log m)=0.086 \cdot \exp\left(-\frac{(\log m -\log 0.22)^2}{2 \cdot 0.57^2}  \right),
\end{equation}

for $m < 0.65~M_\odot$,

\begin{equation}
\label{eq:chabrier_IMF_power}
\xi(m)= c \cdot m^\alpha,
\end{equation}

for $0.65~M_\odot < m$.

The breaking point \citep[where $\xi(m)$ does not follow a power law anymore,][]{Chabrier_03b,Chabrier_03a} is defined at $m=1~M_\odot$ based on empirical fits to observations. To better fit our data, we use as breaking point $m_{50}=0.65~M_\odot$. The normalization is done at the breaking point. The normalization was calculated to fit the slope of the high-mass PDMF determined in the previous section (see Sect.~\ref{sec:mass_funct_slope} and black solid line in Fig.~\ref{fig:MF_FOV}).

To determine the total stellar mass below $M = m_{50}$, we integrated eq.~\ref{eq:chabrier_IMF_log}. For the survey area we determine a total stellar mass of $M_{\rm tot}=(4.4 \pm 0.4) \cdot 10^4~M_\odot$, while the Wd2 cluster mass is estimated to be $M_{\rm Wd2}=(3.6 \pm 0.3) \cdot 10^4~M_\odot$. In Tab.~\ref{tab:masses} we list all masses of the different subregions (as described in Sect.~\ref{sec:spatial_distribution}). Neglecting the unresolved massive binaries causes an additional total mass uncertainty of $\Delta M_{\rm tot}=199~M_\odot$.

\begin{deluxetable}{cccc}
	\tablecaption{The integrated stellar masses \label{tab:masses}}
	\tablehead{
			\multicolumn{1}{c}{Region} &\multicolumn{1}{c}{$m < m_{50}$} & \multicolumn{1}{c}{$m > m_{50}$} & \multicolumn{1}{c}{$m_{tot}$}\\
			\multicolumn{1}{c}{ } &\multicolumn{1}{c}{$[10^4~M_\odot]$} & \multicolumn{1}{c}{$[10^4~M_\odot]$} & \multicolumn{1}{c}{$[10^4~M_\odot]$}
		}
	\startdata
	Survey area 	& $3.9 \pm 0.1$ & $0.5 \pm 0.2$   &	$4.4 \pm 0.3$	\\
	MC 				& $2.0 \pm 0.1$ & $0.3 \pm 0.2$   &	$2.3 \pm 0.2$	\\
	NC 				& $0.3 \pm 0.01$ & $0.03 \pm 0.02$   &	$0.3 \pm 0.03$	\\
	Wd2 			& $3.1 \pm 0.1$ & $0.4 \pm 0.2$   &	$3.6 \pm 0.3$	\\
	Periphery		& $0.4 \pm 0.01$  & $0.1 \pm 0.04$  &	$0.5 \pm 0.05$	\\
	\enddata
	\tablecomments{The integrated stellar masses of the different regions of Wd2. Wd2 is the region within the 2$\sigma$ boundary of the surface density fit (see Sect.~\ref{sec:spatial_distribution}) and periphery is the area outside the same boundary. Column~1 shows the different regions as dedicated in Fig.~\ref{fig:spatial_distribution}. Column~2 gives the total mass for $m < m_{50}$ estimated via the integrated \citet{Chabrier_03a} MF (eq.~\ref{eq:chabrier_IMF_log}). Column~3 gives the total mass for $m > m_{50}$ determined from the observations. Column~4 lists the inferred total stellar mass of the system.}
\end{deluxetable}

Our inferred total stellar mass of $M_{\rm Wd2}=(3.6 \pm 0.3) \cdot 10^4~M_\odot$ is in good agreement with the only mass estimate done so far by \citet{Ascenso_07} who derived a lower limit of $m \ge 10^4~M_\odot$. Our estimate shows that Wd2 is $\sim 75\%$ as massive as the most massive young star cluster in the MW, Wd1 \citep[$M_{\rm Wd1}=4.91^{+1.79}_{-0.49} \cdot 10^4~M_\odot$,][]{Gennaro_11}.

\subsection{Mass segregation in Wd2}
\label{sec:mass_segregation}

\begin{figure}[htb]
	\resizebox{\hsize}{!}{\includegraphics{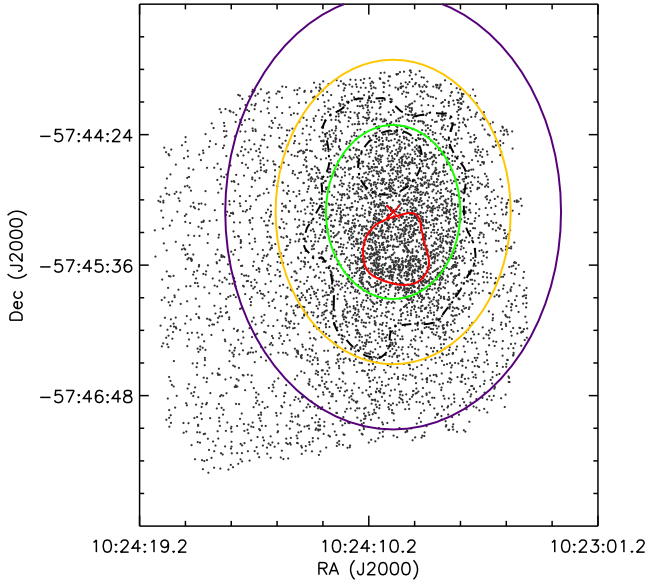}}
	\caption{The different areas used to analyze the mass segregation of Wd2. The midpoint is indicated by the red cross. The different areas are named (from the center outwards) : Wd2, annulus 1, annulus 2, annulus 3, outskirts. The NC and the 2$\sigma$ contours are plotted as reference (dashed contours).}
	\label{fig:spatial_distribution}
\end{figure}

To investigate whether mass segregation has already taken place in the Wd2 cluster we analyze the slope of the PDMF in relation to the distance from the cluster center. As described in Sect.~\ref{sec:spatial_distribution} Wd2 is characterized by a double-peaked, elliptical stellar distribution, thus the usual approach of determining the slope of the PDMF in annuli around the cluster center \citep[e.g.,][]{Sagar_88,Sabbi_08,Parker_15} is not applicable. Instead we chose to use elliptically shaped areas. We defined the midpoint (R.A.=$10^{\rm h}24^{\rm m}01^{\rm s}.94$ and decl.=$-57^\circ45'06.7''$) between the centers of the two clumps as the center of the ellipses. To avoid confusion we will call this center ``midpoint''. As the innermost region we used the MC with its peak coordinates as origin (red contour in Fig.~\ref{fig:spatial_distribution}). We also tried to use the NC but the lower number of stars, especially the low number of high-mass MS stars made it impossible to determine a slope for the MF defined by PMS and MS stars. The next further outwards lying region is called ``annulus~1'' (green contour in Fig.~\ref{fig:spatial_distribution}) encircling both clumps (MC and NC), but we need to note here that this area excludes the MC and NC stars (defined as the $1\sigma$ contours of the stellar density distribution, see Sect.~\ref{sec:spatial_distribution}). The next two larger annuli (``annulus~2'' and ``annulus~3') are shown in yellow and purple contours, while the remaining region around the second annulus we call the ``outskirts''. So in total we divided the survey area in 5 regions from the midpoint outwards.

\begin{figure}[htb]
	\resizebox{\hsize}{!}{\includegraphics{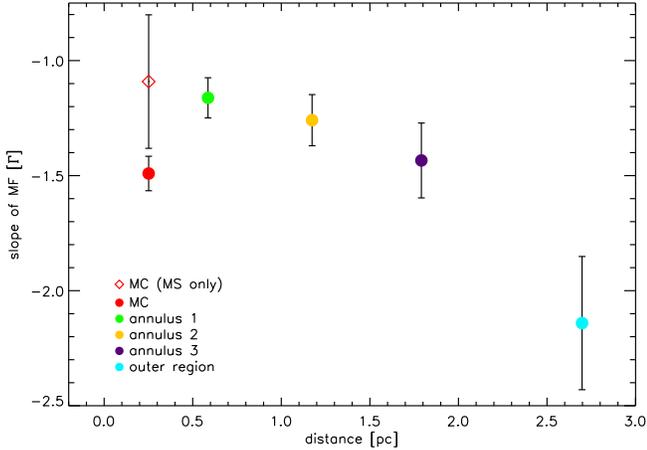}}
	\caption{The slope of the MF for the different regions, plotted against the distance from the center (summary of the data is presented in Tab.~\ref{tab:mass_seg}). The color scheme is the same as in Fig.~\ref{fig:spatial_distribution}. The open diamond represents the MS slope for the MC.}
	\label{fig:MF_slopes}
\end{figure}

We determined the PDMF for each of the different areas. In Tab.~\ref{tab:mass_seg} we list the mean stellar distance for each of the individual regions to the midpoint and the PDMF slope. In Fig.~\ref{fig:MF_slopes} we show the change of the PDMF slope with increasing radial distance from the midpoint. A steeper slope implies many more low-mass stars than high-mass stars.

\begin{deluxetable}{ccc}
	\tablecaption{The slope of the MF relative to the cluster center \label{tab:mass_seg}}
	\tablehead{
			\multicolumn{1}{c}{region} & \multicolumn{1}{c}{mean dist. [pc]} &\multicolumn{1}{c}{$\Gamma$}
		}
	\startdata
	MC 			 & 0.25\tablenotemark{a}	&	$-1.49 \pm 0.07$	\\
	MC (MS only) & 0.25\tablenotemark{a}	&	$-1.09 \pm 0.29$	\\
	annulus 1 	& 0.59	& 	$-1.16 \pm 0.09 $ 	\\
	annulus 2 	& 1.32 	&	$-1.26 \pm 0.11 $ 	\\
	annulus 3	& 1.94 	&	$-1.43 \pm 0.16 $ 	\\
	ouskirts	& 2.70 	&	$-2.14 \pm 0.29 $ 	\\
	\enddata
	\tablecomments{The slopes of the MF for the different regions with increasing distance to the midpoint. In column~2 we give the mean distance from the midpoint. Column~3 shows the slope of the MF including the combined observational uncertainties. \tablenotetext{a}{The mean distance to the MC peak position}}
\end{deluxetable}

As one can see in Fig.~\ref{fig:MF_slopes} and Tab.~\ref{tab:mass_seg}, the slope of the MF gets gradually steeper while moving outwards, away from the cluster center. This means that in the cluster center, there are more massive stars in relation to the lower-mass stars. This implies mass segregation in the cluster center, as is seen in many young massive star clusters like NGC~3603 \citep{Pang_13} or NGC~346 \citep{Sabbi_08}.

It is still a matter of debate whether mass segregation in such young clusters is primordial, meaning that the more massive stars formed originally in the central regions of the cluster, or whether it is dynamical, implying that they formed equally throughout the region and moved inwards due to interactions with the numerous low-mass stars. To evaluate which of the two scenarios is more probable, we will compare the mass segregation time scale with the mean age of the cluster of $\sim1$~Myr (Paper~I).

The mass segregation time scale, $t_{\rm{msg}}$, is defined by:

\begin{equation}
\label{eq:t_msg}
t_{\rm{msg}}=2\cdot t_{\rm{relax}} \cdot \frac{m_{\rm{av}}}{m_{\rm{max}}},
\end{equation}

where $t_{\rm{relax}}$ is the relaxation time of the cluster \citep{Spitzer_69}, $m_{\rm{av}}$ the average stellar mass and $m_{\rm{max}}$ the mass of the most massive star in the cluster \citep{Kroupa_04}. The relaxation time $t_{\rm{relax}}$ depends on the crossing time:

\begin{equation}
\label{eq:t_relax}
t_{\rm{relax}}=\frac{N}{8 \ln N} \cdot t_{\rm{cross}},
\end{equation}

where $t_{\rm{cross}}$ is the crossing time and $N$ the number of stars in the cluster \citep[e.g.,][]{Binney_87} with $t_{\rm{cross}}=2 \cdot r^{3/2} \cdot (G M)^{-1/2}$. Here, $G$ is the gravitational constant and $r$ and $M$ are the total radius and mass of the cluster, respectively.

To estimate the mass-segregation time scale we used all stars enclosed by the annulus~2 (enclosing the Wd2 cluster; see Fig.~\ref{fig:spatial_distribution}). We calculated an approximate crossing time of $t_{\rm{cross}} \approx 1$~Myr using a radius of $\sim 2$~pc. Combining eq.~\ref{eq:t_msg} and eq.~\ref{eq:t_relax} and setting $m_{\rm{max}}=80~M_\odot$ based on the most massive star known in Wd2, we find a mass segregation time scale of $t_{\rm{msg}} \ge 3$~Myr, which is at least three times the mean age of the cluster. This time-scale estimate is a lower limit to the true time scale for mass segregation, since we have not corrected for unresolved binaries, which would increase the number of stars and, as a result, the relaxation time (compare with eq.~\ref{eq:t_relax}).  We can therefore conclude that the mass segregation is, most likely, primordial. We will discuss this in detail in the next Section.

\section{Discussion and Conclusions}
\label{sec:summary}

In this paper we analyzed and discussed the spatial distribution of Wd2's stellar population using our recent optical and near-infrared HST dataset (see Paper~I) and derived the PDMF throughout the cluster area. Combining the determined PDMF with a \citet{Chabrier_03a} IMF to fit the low-mass and brown-dwarf regime, we were able to estimate the integrated stellar mass of the Wd2 cluster.

We used a combination of two 2D Gaussians with a common offset to fit the spatial distribution (see Sect.~\ref{sec:dist}) of the completeness-corrected cluster members and improved the 1D method of Paper~I. For the peak coordinates of the MC we got R.A.=$10^{\rm h}24^{\rm m}01^{\rm s}.65$ and decl.=$-57^\circ45'33.4''$ (J2000) with a peak density of $\Sigma=1863$~stars~pc$^{-2}$. For the NC we got R.A.=$10^{\rm h}24^{\rm m}02^{\rm s}.16$ and decl.=$-57^\circ44'39.3''$ (J2000) with a peak density of $\Sigma=937$~stars~pc$^{-2}$. We defined the sizes of the two clumps as the $1\sigma$ width of the Gaussian density distributions and the area of the Wd2 cluster as the $2\sigma$ width. This led to sizes of 0.31~arcmin$^{2}$ (0.44~pc$^2$) for the MC and 0.27~arcmin$^{2}$ (0.38~pc$^2$) for the NC.

Based on the artificial star tests we carried out a detailed analysis of the PDMF of the Wd2 cluster region (see Sect.~\ref{sec:mass_function}). Only one PDMF has been published for Wd2 before by \citet{Ascenso_07}, using ground-based photometry in the NIR ($J$, $H$, $K_s$). The \citet{Ascenso_07} PDMF is based on a mass rage of 0.8--25$~M_\odot$, representing their completeness limits (detection and saturation limits). They used a distance of 2.8~kpc for Wd2 leading to an underestimation of the stellar masses as compared to our study \citep{Zeidler_15}. Because of the larger wavelength coverage of our data we can better separate the foreground population from the cluster population and due to the higher depth we cover a larger mass range, leading to a better determined PDMF.

We used the photometric catalog of \citet{Vargas_Alvarez_13}, obtained with WFPC2, and spectroscopic data of \citet{Rauw_05,Rauw_07,Rauw_11} and \citet{Vargas_Alvarez_13} to complement our photometric catalog (Paper~I) at the high-mass end with the most luminous stars that are saturated in our data. Using the PARSEC 1.2S isochrones and evolutionary tracks of \citet{Bressan_12} and \citet{Bressan_13} we estimated masses for all of our cluster members. Binning these stellar masses and correcting the number of stars in each mass bin by the corresponding completeness, we were able to estimate a slope of $\Gamma=-1.53 \pm 0.05$ for the PDMF of the survey area (see Fig.~\ref{fig:MF_FOV}). For the Wd2 cluster we estimated a slope of $\Gamma=-1.46 \pm 0.06$. This slope is steeper than the canonical \citet{Salpeter_55} slope of $\Gamma=-1.35$ or \citet{Kroupa_01} slope of $\Gamma=-1.3 \pm 0.3$ but agrees with both within the uncertainties. The slope is also steeper than the slope derived by \citet{Ascenso_07} of $\Gamma=-1.2 \pm 0.16$, based on NIR photometry. To take into account the completeness limit of the ground-based data of \citet{Ascenso_07}, which is $\sim 1.8\,{\rm M}_\odot$ and the different adopted distance we compare the slope of the MS ($\Gamma_{\rm MS}=-1.01 \pm 0.22$) derived in this work with the \citet{Ascenso_07} one. These slopes agree.

The PDMF of Wd2 also agrees very well with other young massive clusters, such as Westerlund~1, the most massive young Galactic star cluster ($m > 4.91^{+1.79}_{-0.49} \cdot 10^4~\rm{M}_\odot$) with an age of $\sim4$~Myr. \citet[][and references therein]{Gennaro_11} determined a slope of $\Gamma=-1.44^{+0.56}_{-0.08}$ (for stars with $m > 3.5~\rm{M}_\odot$). \citet{Cignoni_09} estimated a slope of $\Gamma=-1.25 \pm 0.22$ for NGC~602 (mass range: 0.7--30~M$_\odot$, age: $\sim5$~Myr) and \citet{Sabbi_08} a slope of $\Gamma=-1.87 \pm 0.41$ for NGC~346 (mass range: 8--60~M$_\odot$, age: $\sim3$~Myr), both located in the SMC.

\citet{Weisz_15} studied the MFs of 85 resolved, young (4~Myr~$< t < 25$~Myr) star clusters ($10^3$--$10^4~M_\odot$) in M31 as part of the Panchromatic Hubble Andromeda Treasury program \citep[PHAT,][]{Dalcanton_12} and found a quite universal slope of $\Gamma=-1.45^{+0.03}_{-0.06}$ with little variance for stellar masses $>1~\rm{M}_\odot$. \citet{Weisz_15} reanalyzed also data collected by \citet{Massey_03} for MW and LMC clusters finding slopes of $\Gamma_{\rm{MW}}=-1.16$ and $\Gamma_{\rm{LMC}}=-1.29$ both with a scatter of $\sigma \sim 0.3$--0.4, in agreement with other MW studies and our result for Wd2.

When we subdivide the survey area into elliptical annuli (see Fig.~\ref{fig:spatial_distribution}), outwards from the midpoint, defined as the mean distance between the MC and NC (R.A.=$10^{\rm h}24^{\rm m}01^{\rm s}.94$ and decl.=$-57^\circ45'06.7''$; see Fig.~\ref{fig:spatial_distribution}) and determine the MF for each of the rings, a steepening of the MF slope with increasing distance from the midpoint can be observed (see Fig.~\ref{fig:MF_slopes} and Table~\ref{tab:mass_seg}). This is a clear indication of a mass-segregated cluster population, which is similarly observed in other young massive clusters such as NGC~3603 \citep{Pang_13} and the Arches cluster \citep{Stolte_02} in the MW or NGC~346 \citep{Sabbi_08} in the SMC.

The most massive star WR20a is located outside the cluster center (projected distance: $d\approx 30 ''$), but if we look at the PDMF of the survey area (see Fig.~\ref{fig:MF_FOV}, the two components of WR20a are located in the most massive mass bin) it fits the PDMF slope, which would be consistent with possible membership of WR20a in the Wd2 cluster. \citet{Rauw_05} and \citet{Carraro_13}  discussed the location of WR20a. Both groups argue that WR20a is most likely a member of Wd2 despite being odd that the two most massive member stars of Wd2 are located outside the cluster center. One explanation for its displacement may be that turbulence in the molecular cloud led to multiple sites of star formation and WR20a actually formed at its current location, although so far there has been no evidence of an overdensity of a low-mass PMS population in this region. Another possibility would be that WR20a actually formed in the cluster center and got ejected. \citet{Rauw_05} argue that the rather low velocity of $\sim 0.5$~km~s$^{-1}$ (based on its current position and age) would not be consistent with the ejection hypothesis, yet on the other hand, the initial ejection velocity could have been larger, depending on the cluster potential.

To answer the question whether the mass segregation is, due to the young age of Wd2, primordial or whether it is caused by two-body relaxation we compared the mean cluster age of 1~Myr with an estimate of the mass segregation time scale ($t_{\rm{msg}}$, see Sect.~\ref{sec:mass_segregation}). The mass segregation timescale $t_{\rm{msg}}$ was estimated using eq.~\ref{eq:t_msg} with the size of the cluster, its mean stellar mass estimated using the PDMF, and the most massive star detected in Wd2. The estimated  lower limit of the mass segregation time scale of 3~Myr indicates that the mass segregation is largely primordial \citep[e.g.,][]{Bonnell_06} and not caused by an inward migration of massive stars due to interactions with the large low-mass star population.

Primordial mass segregation is also supported by the work of \citet{Haghi_15} on N-body simulations of star clusters. They argue that primordial mass segregation and mass segregation due to dynamical evolution influence the shape of the PDMF slope in the low-mass regime ($m \le 0.5~\rm{M}_\odot$) differently. In their simulations \citet{Haghi_15} see a significant flattening of the mass function in the low-mass regime similar to what we detect (see Fig.~\ref{fig:MF_FOV}), yet the very low completeness of our data in the center of the two clumps (MC and NC) due to crowding (see Sect.~\ref{sec:completeness}) makes it hard to obtain reliable results.

On the other hand e.g., \citet{McMillan_07} and \citet{Moeckel_09b} argue that the process of merging subclusters keeps the mass segregation imprint of these individual subclusters. We tried to determine the degree of mass segregation by using elliptical annuli around the center of each of the clumps. Due to the low number of stars we could not determine a statistically significant result. Because of the smaller size and lower mass of the subclusters it is difficult to determine whether the mass segregation in each individual clump is primordial or caused by relaxation. The mass segregation time scale of subclusters is much shorter than the one for Wd2 ($t_{\rm{msg}}^{\rm{MC}} \approx 0.1$~Myr and $t_{\rm{msg}}^{\rm{NC}} \approx 0.1$~Myr). The results of \citet{McMillan_07} and \citet{Moeckel_09b} led to the conclusion that due to the clumpiness of the stellar distribution in Wd2 it is difficult to determine if the mass segregation is primordial or caused by interactions.

To complicate things even further, \cite{Allison_10} argued that mass segregation can be introduced dynamically through violent relaxation much more rapidly than through two-body relaxation. Taking into account all these different scenarios, the origin of mass segregation in Wd2 is most likely a combination of different mechanisms (see also \citet{Moeckel_09b}). Nevertheless, due to the cluster's very young age and high degree of mass segregation, we favor the scenario of primordial mass segregation to be the main cause of the observed mass segregation in Wd2.

Adopting the \citet{Chabrier_03a} IMF (see eq.~\ref{eq:chabrier_IMF_log}) for the low-mass regime, where the completeness limits drop below 50\%, we estimate a total stellar mass for the Wd2 cluster of $(3.6 \pm 0.3) \cdot 10^4~M_\odot$, which is in good agreement with the first estimate of \citet{Ascenso_07} of $M \ge 10^4~M_\odot$. With this mass Wd2 has about 75\% of the stellar mass of Wd1 \citep{Gennaro_11}, the most massive star cluster in the MW.

The comparison of the subcluster surface densities calculated in our study ($\Sigma_{NC}=555$~stars~pc$^{-2}$ and $\Sigma_{MC}=1114$~stars~pc$^{-2}$) with the results of \citet{Kuhn_15a} places the subclusters of Wd2 at the lower end of the surface density range of $\sim1$ to $\sim30,000$~stars~pc$^{-2}$ \citep{Kuhn_15a}. Therefore, this result is comparable with the surface density of, e.g., M~17. We have to note here that we applied the lower mass cut at $0.65~M_\odot$ (\citet{Kuhn_15a} applied a lower mass limit at 0.1--$0.2~M_\odot$) in order to properly correct for completeness effects. Therefore, our values should be considered as lower limits. In addition, \citet{Kuhn_15a} found that there is no special value for the surface density of subclusters.

\citet{Banerjee_15a} explore various scenarios of subclustering for massive ($>10^4$~M$_\odot$) star clusters when studying the possible formation mechanisms of the young massive cluster NGC~3603 \citep[e.g.,][]{Stolte_04,Pang_11}. The proposed scenarios are either in-situ monolithic cluster formation, meaning the cluster forms in a single, massive star-formation event, or via merging of less massive subclusters, which then fall onto each other and merge to form the final massive cluster. \citet{Banerjee_15a} argue that the monolithic cluster formation with an in-situ formation fits best the observations of NGC~3603 \citep{Banerjee_14}, but also the scenario of the cluster assembly from subclusters within 1~Myr is not ruled out. In one simulation they used as initial conditions a number of 10 subclusters, each following a Plummer sphere \citep{Plummer_1911} distribution of gas. The subclusters were equally distributed over a sphere with radius $R=2.5$~pc. After 1~Myr this simulation gives a system with two subclusters, separated by $\sim 1$~pc (cf. their Panel~5 in Fig.~5 and Panel~3 in Fig.~6). Their simulated system is still in process of merging at $t\approx1$~Myr and far from virialization and a final spherical shape. The distance between the two clumps in Wd2 is 54.2~arcsec or 1.08~pc and therefore well comparable with the results of these simulations, while the cavity in RCW~49 cleared by the cluster has a radius of $\sim2$~pc. Due to the similar age of Wd2 and NGC~3603 and the existence of two coeval subclusters that fit well the properties of the simulations in \citet{Banerjee_15a} a formation from subclusters is a likely scenario for the formation of Wd2.

The presence of a faint, uniformly distributed low-mass population ($<0.15M_\odot$), visible throughout the RCW~49 area, implies that the actual star-forming region is more extended than the Wd2 cluster. This phenomenon was already observed by \citet{Whitney_04}, who identified $\sim 300$ cYSOs in the giant \ion{H}{2} region RCW~49 using $Spitzer$ IRAC MIR data of the GLIMPSE survey \citep{Benjamin_03}. \citet{Churchwell_04} described the RCW~49 gas and dust distribution as ``a complex network of thin filaments, pillars, sharp boundaries, and knots''. This irregular distribution and the associated gas, probably created by the feedback from the massive stellar population, is a perfect birthplace for a new population of stars. \citet{Whitney_04} identified 5 more star-forming sites throughout the RCW~49 cloud. This is also observed in other young star clusters, e.g., in NGC~3603, the Orion Nebula Cluster \citep[e.g.,][]{Hillenbrand_97}, or the 30 Doradus region \citep[e.g.,][]{Walborn_99,Brandner_01}.

A possible scenario for the formation of a massive star cluster is the cloud-cloud collision of giant molecular clouds. This can lead to a long-lasting, extended formation of a low-mass stellar population, similar to what we see in our data. Cloud-cloud collision is also considered as one of the modes for high-mass star formation. Theoretical studies and hydrodynamical simulations for the scenario of cloud-cloud collision were carried out by \citet{Habe_92,Anathpindika_10} and \citet{Takahira_14}, among others. This scenario has been suggested by \citet{Nigra_08} and \citet{Cignoni_09} for the star cluster NGC~602 in the SMC and is also debated for other star clusters such as \object{NGC~3603} \citep{Fukui_14} and RCW~38 \citep{Fukui_16} in the MW. For the RCW~49 region this scenario is supported by CO observations by \citet{Furukawa_09,Furukawa_14} and \citet{Ohama_10} using data of the NANTEN survey \citep{Mizuno_04} as well as a large population of young PMS stars detected in the RCW~49 gas cloud \citep{Whitney_04}, far away from the influence of the luminous OB star population.

\citet{Ohama_10,Furukawa_09} and \citet{Furukawa_14} showed that the CO distribution throughout RCW~49 splits up into three different clouds with a LSR velocity of 16, 4, and -4~km~s$^{-1}$, respectively. \citet{Furukawa_09} calculated the respective kinematic distances to be 6.5, 5.2, and 4.0~kpc using the rotation curve of \citet{Brand_93}. In the dereddened $F814W_0$ vs. $(F814W-F160W)_0$ CMD we see a red, low-mass population of stars which we refer to as the red tail. We ruled out that photometric uncertainties are responsible for such a feature. Analyzing the location of these red stars on the sky, an overdensity to the south-west becomes visible (see Fig.~\ref{fig:F814W+redmap}). This overdensity coincides with an area with low $F814W$ and H$\alpha$ fluxes (left panels of Fig.~\ref{fig:F814W+redmap} and Fig.~\ref{fig:Halpha+Spitzer}) and is not correlated with the structure seen in the gas extinction map (right panel of Fig.~\ref{fig:F814W+redmap}). Comparing this area to $Spitzer$ IRAC observations (see right panel of Fig.~\ref{fig:Halpha+Spitzer}) of the GLIMPSE survey \citep{Benjamin_03} a molecular cloud becomes visible. This cloud structure is very well visible in the IRAC-3 [5.8] band containing the PAH $6.2\mu$m C--C stretching mode \citep{Draine_03}. This leads to the conclusion that this molecular cloud is located in front of the Wd2 cluster. This conclusion is supported by the fact that we cannot see the cloud in our H$\alpha$ and Pa$\beta$ line emissions and therefore we are underestimating the gas extinction $E(B-V)_g$.

The CO cloud located at a distance of 4.0~kpc \citep{Furukawa_09,Furukawa_14,Ohama_10} coincides with the location of the molecular cloud seen in the $Spitzer$ data and the location of the red-tail members. The cloud kinematics estimated by \citet{Dame_07} and \citet{Furukawa_09} suggest that Wd2 is located behind the 4, and $-4$~km~s$^{-1}$ clouds, which is in contradiction to our distance estimate. On the other hand, if we underestimated the extinction toward Wd2 our distance modulus is overestimated leading to an even closer distance. We suggest that Wd2 is located between the 4 and $-4$~km~s$^{-1}$ clouds in agreement with our distance of 4.16~kpc determined in Paper I.

\acknowledgements
We want to thank Michele Cignoni for fruitful discussions about the PMS models and the artificial star tests.

Based on observations made with the NASA/ESA Hubble Space Telescope, obtained at the Space Telescope Science Institute, which is operated by the Association of Universities for Research in Astronomy, Inc., under NASA contract NAS 5-26555. These observations are associated with program \#13038.

P.Z., E.K.G., and A.P. acknowledge support by Sonderforschungsbereich 881 (SFB 881, ``The Milky Way System'') of the German Research Foundation, particularly via subproject B5. M.T. has been partially funded by PRIN-MIUR 2010LY5N2T.

We thank ESA for the financial support for P.Z. to visit STScI for a productive scientific collaboration.

We thank the referee for the helpful comments to improve the quality of the paper.

\facility{HST (ACS, WFC3)}

\appendix
\section{Tables}

\onecolumngrid
\floattable
\begin{deluxetable}{lrlrlrlrlrlrl}
	\tablecaption{The completeness limits for different regions\label{tab:completeness}}
	\tablehead{
		\multicolumn{1}{c}{ } & \multicolumn{4}{c}{\textsc{complete survey area}}  &\multicolumn{2}{c}{\textsc{ gas ridge\tablenotemark{a}}}&\multicolumn{2}{c}{\textsc{ Wd2\tablenotemark{b}}}&\multicolumn{2}{c}{\textsc{ MC}}&\multicolumn{2}{c}{\textsc{NC}}\\
		\multicolumn{1}{c}{\textsc{Filter}} &\multicolumn{1}{c}{\textsc{50\%}} &\multicolumn{1}{c}{\textsc{M [$M_\odot$]}} &\multicolumn{1}{c}{\textsc{75\%}} &\multicolumn{1}{c}{\textsc{M [$M_\odot$]}} &\multicolumn{1}{c}{\textsc{50\%}} &\multicolumn{1}{c}{\textsc{M [$M_\odot$]}} &\multicolumn{1}{c}{\textsc{50\%}} &\multicolumn{1}{c}{\textsc{M [$M_\odot$]}} &\multicolumn{1}{c}{\textsc{50\%}} &\multicolumn{1}{c}{\textsc{M [$M_\odot$]}} &\multicolumn{1}{c}{\textsc{50\%}} &\multicolumn{1}{c}{\textsc{M [$M_\odot$]}}		
	}
	\startdata
	$F555W$	& 24.8 & 0.70 & 23.6 & 1.10 & 26.2 & 0.52 & 26.6 & 0.49 & 24.3 & 0.80 & 26.5 & 0.41 \\
	$F814W$	&  23.3 & 0.20 & 21.8 & 0.45 & 24.7 & 0.12 & 25.3 & 0.10 & 22.8 & 0.26 & 24.6 & 0.12 \\
	$F125W$	&  20.2 & 0.12 & 18.5 & 0.48 & 21.7 & $<0.1$ & 22.0 & $<0.1$ & 18.7 & 0.44 & 21.2 & $<0.1$ \\
	$F160W$	&  19.4 & 0.12 & 17.8 & 0.45 & 20.9 & $<0.1$ & 20.9 & $<0.1$ & 17.6 & 0.46 & 20.2 & $<0.1$ \\
	$E(B-V)$&  \multicolumn{4}{c}{1.55~mag } & \multicolumn{2}{c}{1.60~mag} & \multicolumn{2}{c}{1.65~mag} & \multicolumn{2}{c}{1.53~mag} & \multicolumn{2}{c}{1.49~mag} \\
	\enddata
	\tablecomments{In this table we list the mean magnitudes and the correspondig stellar masses for various region for a completeness limit of 50\%. The stellar masses correspond to a stellar age of 1~Myr. The $E(B-V)$ color excess is used to obtain masses from the isochrones and is obtained from our high-resolution, 2D color excess map (Paper~I).}
	\tablenotetext{a}{As the gas ridge we refer to the \ion{H}{2} region of RCW~49 toward the south east of the cluster.}
	\tablenotetext{b}{As the area of Wd2 we refer to the region within the 2$\sigma$ boundary of the spatial distribution (see Sect.~\ref{sec:spatial_distribution}).}
\end{deluxetable}
\twocolumngrid

\bibliographystyle{aasjournal}
\bibliography{../../bibliography/Wd2_bibliography}

\end{document}